\PassOptionsToPackage{pdfpagelabels=false}{hyperref}
\documentclass[fleqn,usenatbib]{mnras}
\usepackage{newtxtext,newtxmath}

\usepackage[T1]{fontenc}
\usepackage{ae,aecompl}
\pdfoutput=1

\usepackage{graphicx}	
\usepackage{amsmath}	
\usepackage{amssymb}	
\usepackage{subfigure}

\newcommand{\FIREurl}{\href{http://fire.northwestern.edu}
{\url{http://fire.northwestern.edu}}}
\newcommand{\gizmourl}{\href{http://www.tapir.caltech.edu/~phopkins/Site/GIZMO.html}{\url{http://www.tapir.caltech.edu/~phopkins/Site/GIZMO.html}}}
\newcommand\altaffilmark[1]{$^{#1}$}
\newcommand\altaffiltext[1]{$^{#1}$}

\defcitealias{ElBadry_2018}{E18}
\defcitealias{Bradford_2015}{B15}

\title[Gas Kinematics from Unresolved HI]{Gas Kinematics in FIRE Simulated Galaxies Compared to Spatially Unresolved HI Observations}

\author[El-Badry et al.]{
\parbox[t]{\textwidth}{ 
Kareem El-Badry\thanks{E-mail: kelbadry@berkeley.edu}\altaffilmark{1},
Jeremy Bradford\altaffilmark{2},
Eliot Quataert\altaffilmark{1}, 
Marla Geha\altaffilmark{2},
Michael Boylan-Kolchin\altaffilmark{3},
Daniel R. Weisz\altaffilmark{1},
Andrew Wetzel\altaffilmark{4}, 
Philip F.~Hopkins\altaffilmark{5},
T.~K.\ Chan\altaffilmark{6}, 
Alex Fitts\altaffilmark{3}, 
Du\v{s}an Kere\v{s}\altaffilmark{6}, 
Claude-Andr{\'e} Faucher-Gigu{\`e}re\altaffilmark{7}
} 
\vspace*{6pt} \\
\altaffiltext{1}{Department of Astronomy and Theoretical Astrophysics Center, University of California Berkeley, Berkeley, CA 94720} \\
\altaffiltext{2}{Department of Astronomy, Yale University, New Haven, CT 06520, USA} \\
\altaffiltext{3}{Department of Astronomy, The University of Texas at Austin, Austin, TX 78712} \\
\altaffiltext{4}{Department of Physics, University of California, Davis, CA 95616} \\
\altaffiltext{5}{TAPIR, Mailcode 350-17, California Institute of Technology, Pasadena, CA 91125} \\
\altaffiltext{6}{Department of Physics, Center for Astrophysics and Space Sciences, University of California at San Diego, La Jolla, CA 92093} \\ 
\altaffiltext{7}{Department of Physics and Astronomy and CIERA, Northwestern University, Evanston, IL 60208} \\ 
}

\date{Submitted to MNRAS, Jan 11, 2018}

\pubyear{2018}

\begin{document}
\label{firstpage}
\pagerange{\pageref{firstpage}--\pageref{lastpage}}
\maketitle

\begin{abstract}
The shape of a galaxy's spatially unresolved, globally integrated 21-cm emission line depends on its internal gas kinematics: galaxies with rotation-supported gas disks produce double-horned profiles with steep wings, while galaxies with dispersion-supported gas produce Gaussian-like profiles with sloped wings. Using mock observations of simulated galaxies from the FIRE project, we show that one can therefore constrain a galaxy's gas kinematics from its unresolved 21-cm line profile. In particular, we find that the kurtosis of the 21-cm line increases with decreasing $V/\sigma$, and that this trend is robust across a wide range of masses, signal-to-noise ratios, and inclinations. We then quantify the shapes of 21-cm line profiles from a morphologically unbiased sample of $\sim$2000 low-redshift, HI-detected galaxies with $M_{\rm star} = 10^{7-11} M_{\odot}$ and compare to the simulated galaxies. At $M_{\rm star} \gtrsim 10^{10} M_{\odot}$, both the observed and simulated galaxies produce double-horned profiles with low kurtosis and steep wings, consistent with rotation-supported disks. Both the observed and simulated line profiles become more Gaussian-like (higher kurtosis and less-steep wings) at lower masses, indicating increased dispersion support. However, the simulated galaxies transition from rotation to dispersion support more strongly: at $M_{\rm star} = 10^{8-10}M_{\odot}$, most of the simulations produce more Gaussian-like profiles than typical observed galaxies with similar mass, indicating that gas in the low-mass simulated galaxies is, on average, overly dispersion-supported. Most of the lower-mass simulated galaxies also have somewhat lower gas fractions than the median of the observed population. The simulations nevertheless reproduce the observed line-width baryonic Tully-Fisher relation, which is insensitive to rotation vs. dispersion support. 
\end{abstract}



\section{Introduction}
The shape of a galaxy's globally integrated 21-cm line profile encodes information about the kinematics of neutral hydrogen (HI) in its interstellar medium (ISM). Because the 21-cm transition is usually optically thin, an observed line profile can be directly translated into a line-of-sight velocity distribution (LOSVD) of HI gas in the galaxy. The shape of the 21-cm line thus depends on a galaxy's internal kinematics, and it will in general be different for disky, rotation-supported galaxies than for irregular galaxies in which gas is primarily supported by dispersion. 

Most large spiral galaxies produce characteristic ``double-horned'' unresolved 21-cm line profiles, which arise naturally from differential Doppler effects in rotating disks \citep{Epstein_1964, Shostak_1977}. In fact, double-horned line profiles are not unique to 21-cm emission from HI in galaxies but are a generic consequence of rotating disk geometries and are produced in a wide range of astrophysical settings, including lines formed in accretion disks in some broad-line AGN \citep{Stella_1990, Storchi_2017}, protostellar disks \citep{Yamada_2009}, and accretion disks around massive stars \citep{Doazan_1965} and white dwarfs \citep{Tylenda_1981}. In many cases, the precise shape of such double-peaked emission lines can be used to constrain the rotation curve and/or spatial distribution of the emitting material. 

Unlike large spiral galaxies, many low-mass galaxies do not produce double-horned profiles, instead exhibiting single-peaked, Gaussian-like profiles with sloping wings \citep[e.g.][]{Geha_2006, Begum_2008, Haynes_2011, Bradford_2015}.  Single-peaked, Gaussian-like profiles are typically interpreted as evidence of lower rotation velocities and more disordered gas kinematics than is typical for more massive disk galaxies \citep{Singhal_2008, Stewart_2014, Papastergis_2016}. This suggests that in the absence of spatially resolved observations, the shape of a galaxy's unresolved HI profile can be used as a rough proxy for $V/\sigma$ (i.e., the ratio of ordered rotation to dispersion) of the galaxy's gas.

In contrast to resolved HI kinematic maps, which are expensive to obtain and have only been produced for of order one hundred galaxies in the local Universe \citep[e.g.][]{Walter_2008, Hunter_2012, Ott_2012}, unresolved global 21-cm line profiles can be efficiently obtained with short, single-dish observations. Unresolved HI profiles are already cataloged for a few $\times\, 10^4$ low-redshift galaxies \citep[e.g.][]{Paturel_2003, Meyer_2004, Begum_2008, Haynes_2011, Giovanelli_2015}, and next-generation radio surveys are expected to increase the size of unresolved HI samples by a factor of 10 within a decade \citep{Dewdney_2009, Duffy_2012, Li_2017}. 

Besides their availability, another advantage of using unresolved 21-cm line profiles as probes of HI kinematics is that most unresolved surveys are morphologically blind, providing a homogeneous and unbiased sample of the local gas-rich galaxy population against which to test the predictions of theoretical models. This is in contrast to spatially resolved studies, which often preferentially target ``well-behaved'' disk galaxies. Mock unresolved HI observations are also straightforward to produce from simulations with minimal post-processing and can be compared directly to observations, without the need to fit kinematic models.

Both observational \citep{McGaugh_2000, Haynes_2011, Bradford_2015, Bradford_2016} and theoretical \citep{Obreschkow_2009, Obreschkow_2009b, Brook_2016, Maccio_2016, Brooks_2017} works have previously interpreted HI profiles from unresolved data. However, these works have typically not used the \textit{shape} of line profiles to extract kinematic information, but have instead quantified observed 21-cm lines with only two numbers: the integrated flux, which can be translated to total HI mass, and the line width, which to first order traces the depth of the gravitational potential. These parameters are not sensitive to whether the emitting gas is supported primarily by rotation or dispersion. 

In this paper, we compare unresolved 21-cm line profiles from simulations to profiles of real galaxies, with the goal of comparing the simulated and observed galaxies' HI kinematics. We first show how the information encoded in the shapes of unresolved 21-cm line profiles can be leveraged to statistically constrain the gas kinematics of the low-redshift galaxy population. We use mock HI profiles produced from simulated galaxies, for which the internal gas kinematics are known a priori, as a guide for interpreting the line profiles of real galaxies. We then compare the line profiles of real galaxies to the mock profiles in order to compare the degree of rotation vs. dispersion support in the simulated galaxies to that in the observed population. 

The remainder of this paper is organized as follows. In Section~\ref{sec:mock_obs}, we describe our procedure for mock-observing simulated galaxies and show how unresolved HI line shape is related to internal rotation vs. dispersion support. In Section~\ref{sec:obs_comp}, we quantify the shapes of observed HI line profiles and compare to our simulated galaxies. We summarize our results and discuss their implications in Section~\ref{sec:conclusion}. In Appendix~\ref{sec:toy_model}, we describe a simple analytic model for predicting the shape of unresolved HI line profiles. We discuss the relation between rotation/dispersion support and other galaxy properties in Appendix~\ref{sec:sfhs}.

\section{Mock observations}
\label{sec:mock_obs}

\begin{figure*}
\centering 
\subfigure[simulated galaxies]{\label{fig:schematic}\includegraphics[width=\columnwidth]{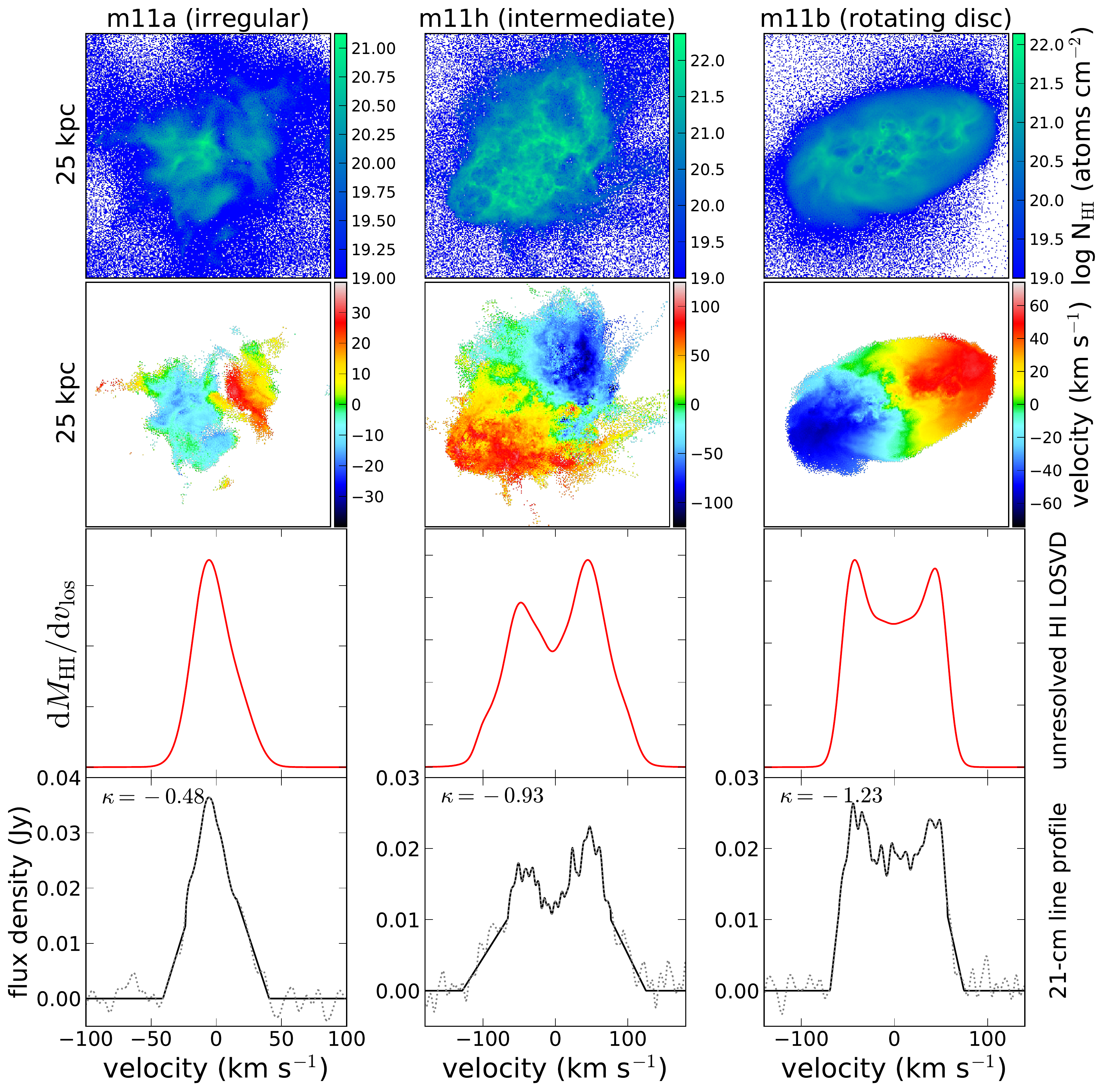}}
\subfigure[real galaxies]{\label{fig:data_schematic}\includegraphics[width=\columnwidth]{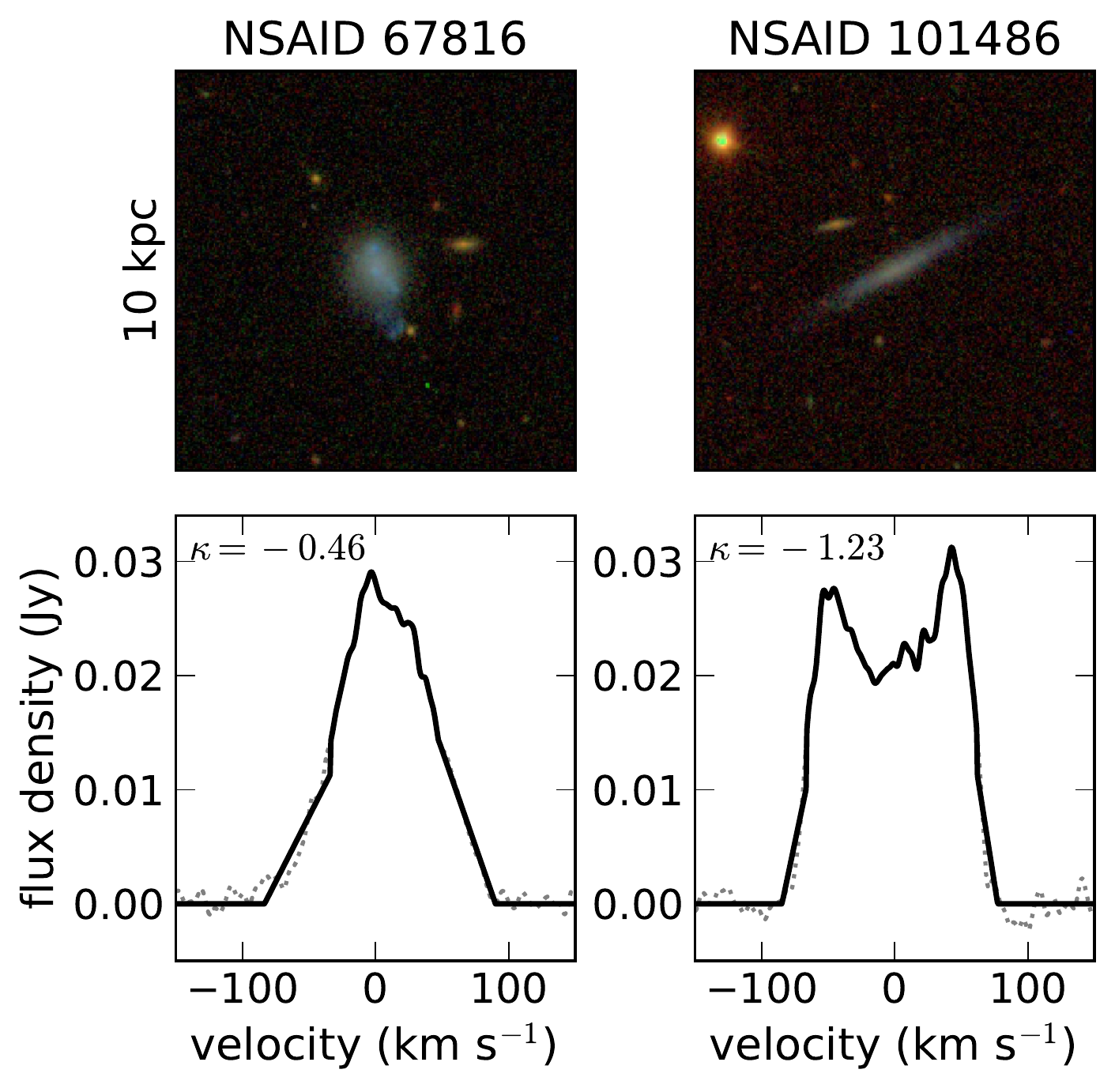}}
\caption{\textbf{Left}: Effect of internal gas kinematics on the shape of galaxies' unresolved 21-cm lines. First (second) row shows the resolved zeroth (first) HI moment map for three low-mass galaxies in our sample, with increasing rotational support from left to right. Each map is produced with inclination $i=60$ deg. Third row shows the global HI mass-weighted line-of-sight velocity distribution. Bottom row shows the mock 21-cm spectrum (gray) and identified emission region (black; see Section~\ref{sec:mock_obs}). Unresolved spectra encode information about the galaxies' gas kinematics: rotation-supported galaxies produce double-horned line profiles with steep wings, while dispersion-supported galaxies produce single-peaked profiles with shallower wings. \textbf{Right}: Bottom panels show unresolved HI data for two low-mass galaxies ($M_{\rm star}\sim 10^8 M_{\odot}$), one with a Gaussian-like profile and one with a double-horned profile. Top panels show corresponding SDSS images.}
\end{figure*}

We analyze the same suite of cosmological hydrodynamic zoom-in simulations from the FIRE project \citep{Hopkins_2014} that was studied by \citet[][hereafter E18]{ElBadry_2018}. The simulations were run with the \texttt{GIZMO} hydrodynamics code \citep{Hopkins_2015} in the Lagrangian ``meshless finite mass'' (MFM) mode, using the FIRE-2 model for galaxy formation and feedback\footnote{See the FIRE project website at \FIREurl. A public version of the \texttt{GIZMO} code is available at \gizmourl.} \citep{Hopkins_2017}. For details regarding the physical processes modeled in these simulations and their numerical implementation, we refer the reader to \citet{Hopkins_2017} and \citet{Hopkins_2017b}.

This simulation suite contains 24 isolated galaxies, with $z=0$ stellar masses spanning $M_{\rm star}=10^{6.3-11.1} M_{\odot}$, residing in halos with $z=0$ masses $M_{\rm 200m} = 10^{9.9-12.2}M_{\odot}$. Here $M_{\rm 200m}$ is the total mass within $R_{\rm 200m}$, which is the radius within which the matter density is 200 $\times$ the mean matter density at $z=0$. \citetalias{ElBadry_2018} studied the gas properties of these galaxies in detail, showing that they exhibit a diverse range of gas kinematics and morphologies: thin, rotation-supported disk galaxies dominate at $M_{\rm star} \gtrsim 10^{10} M_{\odot}$, while puffy, dispersion-supported irregular galaxies dominate at lower masses. The properties of the individual simulated galaxies studied in this work are summarized in \citetalias{ElBadry_2018} (their Table 1). 

We mock-observe these simulated galaxies in order to produce synthetic unresolved HI profiles that can be compared to observations. The observational sample to which we compare is the homogeneously reduced sample of unresolved HI observations produced by \citet[][hereafter B15]{Bradford_2015}; our mock observations are designed to emulate these data. Most of the lower-mass galaxies ($M_{\rm star} \lesssim 10^{8.5} M_{\odot}$) in the observational sample were observed with a typical velocity resolution of $\Delta v = 0.65\,{\rm km\,s^{-1}}$ as part of the low-mass HI program introduced in \citet{Geha_2006}; see Figure 3 of \citetalias{Bradford_2015}. Most of the higher-mass galaxies were originally observed with a velocity resolution of $\Delta v = 5.5\,{\rm km\,s^{-1}}$ through the ALFALFA survey \citep{Haynes_2011} and subsequently re-reduced by \citetalias{Bradford_2015}. Spectra collected through the low-mass HI program were Hanning smoothed to a resolution of $5\,{\rm km\,s^{-1}}$ by \citetalias{Bradford_2015}; those from the ALFALFA survey, to a resolution of $10\, {\rm km\,s^{-1}}$. In this work, we only analyze the subsample of the \citetalias{Bradford_2015} data with a smoothed signal-to-noise ratio ${\rm SNR} > 15$ (see Equation~\ref{eqn:snr_eqn}); this subsample contains 2002 galaxies with a median SNR of 25.  

For our main analysis, we mock-observe all simulated galaxies at $z=0$ from 100 random viewing angles distributed uniformly on the unit sphere. For a given viewing angle and desired signal-to-noise ratio, we apply the following procedure to create a mock-HI profile:

\begin{enumerate}
\item We consider all HI gas within a circular aperture of diameter 80 kpc centered on the galaxy. This is comparable to the 3.5 arcmin Arecibo beam diameter for an object at a distance of 70 Mpc, which is typical for our sample of observations. Our results are not sensitive to this choice, because the aperture is larger than the region containing significant HI for all the simulated galaxies in our sample. We calculate the HI mass of each gas element in the simulation as in \citetalias{ElBadry_2018}, excluding ionized gas and cold gas likely to be molecular.\footnote{Gas with $T<300\,\rm K$ and $n_{\rm H} > 10\,{\rm cm^{-3}}$ is treated as molecular. Our results are not sensitive to excluding molecular gas, as only a small fraction ($\sim$10\% by mass in MW-mass galaxies, and less at lower masses) of the ISM passes these molecular criteria at $z=0$. We note that these molecular fractions are likely somewhat lower than typical values for observed galaxies.} The neutral hydrogen fraction of each gas element is computed by \texttt{GIZMO}, which incorporates ionization from a spatially uniform, redshift-dependent UV background \citep{FG_2009} and approximates the radiation field due to local sources (see \citealt[][]{Hopkins_2017}).
\item We construct the HI mass-weighted LOSVD of all HI gas in this aperture, binned at a velocity resolution $\Delta v$. To account for thermal broadening, we treat each gas element as a Gaussian in velocity space with $\sigma = c_s=\sqrt{kT/(\mu m_p)}$, where $c_s$ is the isothermal sound speed. 
We choose $\Delta v$ to match the observations against which we compare: for galaxies with $M_{\rm star} < 10^{8.5} M_{\odot}$, we use $\Delta v = 0.65\,\rm km\,s^{-1}$; for more massive galaxies, we take $\Delta v = 5.5\,\rm km\,s^{-1}$.

\item For mock-observations of a target at a distance $D$, we compute the 21-cm flux density in the $i$-th velocity channel using the formula 
\begin{equation}
\label{eqn:S_nu}
\left(\frac{S_{{21,i}}}{{\rm Jy}}\right)=\frac{1}{2.36\times10^{5}}\left(\frac{m_{{\rm HI},i}}{M_{\odot}}\right)\left(\frac{\Delta v}{{\rm km\,s^{-1}}}\right)^{-1}\left(\frac{D}{{\rm Mpc}}\right)^{-2},
\end{equation}
implicitly assuming the 21-cm transition is optically thin \citep[e.g.][]{Haynes_1984}. Here $m_{{\rm HI},i}$ is the HI mass in the $i$-th velocity channel. In practice, we solve for $D$ for each observation in order to yield the desired signal-to-noise ratio, as quantified in Equation~\ref{eqn:snr_eqn}. 

\item We add Gaussian noise to the spectrum with a fixed dispersion of 0.004 Jy, which is typical for the observations against which we compare. We then Hanning smooth the spectrum to a resolution consistent with the observations: for galaxies with $M_{\rm star} < 10^{8.5} M_{\odot}$, we smooth to a resolution of $5\,\rm km\,s^{-1}$; for higher-mass galaxies, to a resolution of $10\,\rm km\,s^{-1}$.

\item We identify the HI emission region of the spectrum in velocity space as follows. We construct a copy of the noisy spectrum that is Hanning smoothed to twice the resolution adopted in (iv). We define the edges of the emission region as the first point on either side of the emission peak where this doubly-smoothed spectrum drops below zero flux; we set the emission to zero outside this region. 

Denoting the peak flux density $f_{\rm peak}$, we fit a quadratic polynomial to the wings on either side of the line profile between $0.5f_{\rm peak}$ and the edges of the emission region, replacing the flux in this region with the polynomial interpolation. This procedure was found by \citetalias{Bradford_2015} to minimize spurious signal due to noise in the wings of the profile; we adopt it here for consistency. 

\item We calculate the total integrated flux of the HI line as $S_{21,{\rm tot}}=\int_{-\infty}^{\infty}S_{{\rm 21}}\left(v\right)\,{\rm d}v$, and $W_{\rm 50}$, the velocity width at 50\% of the maximum flux, as the difference between the velocities above and below the peak flux at which the interpolated emission profile reaches $0.5 f_{\rm peak}$. We then calculate the signal-to-noise ratio of the integrated emission using the same definition used by \citetalias{Bradford_2015}:
\begin{equation}
\label{eqn:snr_eqn}
{\rm SNR}=\left(\frac{S_{21,{\rm tot}}}{W_{50}}\right)\frac{w_{{\rm smo}}^{1/2}}{\sigma_{{\rm rms}}}.
\end{equation}
Here  $w_{\rm smo}=W_{50}/(10\rm\,km\,s^{-1})$, and $\sigma_{\rm rms}$ is the standard deviation of the spectrum outside the emission region, after the smoothing in (iv) is applied. Note that this definition represents the total signal-to-noise ratio of the HI line integrated over all channels; the SNR in any one channel will be lower than the value yielded by Equation~\ref{eqn:snr_eqn}. 

\end{enumerate}

\subsection{Quantifying Unresolved Line Shapes}
\label{sec:kurt}
We use two distinct but related statistics to quantify the shape of galaxies' integrated HI lines.

\subsubsection{Kurtosis}
The kurtosis, which we denote $\kappa$, is a measure of the ``peakedness'' of the line; i.e., how centrally concentrated the emission is in velocity space. Single-peaked, Gaussian-like HI lines have relatively high kurtosis, while line profiles with more weight in the edges, such as the characteristic double-horned profile, have lower kurtosis. The kurtosis of galaxies' integrated HI lines was previously used as a proxy for $V/\sigma$ by \citet{Papastergis_2016}, who used a cut of $\kappa \leq -1.2$ to select rotation-supported galaxies in a study of the local baryonic Tully-Fisher relation (BTFR). For an observed line profile $S_{21}(v)$, we define the kurtosis as 

\begin{equation}
\label{eqn:kurt}
\kappa=\frac{\mu_{4}}{\mu_{2}^{2}}-3.
\end{equation}
Here $\mu_n$ is the $n$th central moment:
\begin{equation}
\label{eqn:cen_mom}
\mu_{n}=\frac{\int_{-\infty}^{\infty}S_{{\rm 21}}\left(v\right)\left(v-\overline{v}\right)^{n}\,{\rm d}v}{\int_{-\infty}^{\infty}S_{{\rm 21}}\left(v\right)\,{\rm d}v},
\end{equation}
where $\overline{v}$ is the flux-weighted mean velocity:
\begin{equation}
\label{eqn:mean_v}
\overline{v}=\frac{\int_{-\infty}^{\infty}v S_{{\rm 21}}\left(v\right)\,{\rm d}v}{\int_{-\infty}^{\infty}S_{{\rm 21}}\left(v\right)\,{\rm d}v}.
\end{equation}
The constant $-3$ in Equation~\ref{eqn:kurt} sets the kurtosis scale so that a Gaussian has $\kappa=0$. We find that for both our mock observations and real galaxies, the majority of HI lines have $-1.5 \lesssim \kappa \lesssim 0$. 

\subsubsection{Steepness of profile wings}
To quantify the steepness of a profile's wings, we first calculate $W_{20}$ and $W_{50}$, the width of the line profile at 20\% and 50\% of its peak value. We then measure their scaled difference, $\Delta W \equiv (W_{\rm 20} - W_{\rm 50})/W_{50}$. This quantity is a measure of the slope of the line profile wings: for profiles with steep wings, $W_{20}$ and $W_{50}$ are very similar, and $\Delta W$ is small. For profiles with sloping wings, $W_{\rm 20}$ is significantly larger than $W_{50}$, so $\Delta W$ is larger; for a Gaussian, $\Delta W = 0.52$. Most of the line profiles in both our observed and simulated samples have $0 \lesssim \Delta W \lesssim 1$.

\subsection{Relation between line shape and kinematics}

In Figure~\ref{fig:schematic}, we show how a galaxy's gas kinematics change the shape of its unresolved 21-cm profile. We show column density and velocity maps of three simulated galaxies spanning the range of gas kinematics and morphology in our sample, ranging from irregular and dispersion supported (\texttt{m11a}, with $M_{\rm star} = 10^{8}M_{\odot}$) to rotationally supported but with significant disordered motion (\texttt{m11h}, with $M_{\rm star} = 4\times 10^{9}M_{\odot}$), to rotationally supported and morphologically disky (\texttt{m11b}, with $M_{\rm star} = 10^{8}M_{\odot}$). Details regarding the construction of these maps, as well as comparable maps for all the galaxies in our sample, can be found in \citetalias{ElBadry_2018}. Figure~\ref{fig:schematic} shows that the simulated galaxies' unresolved HI profiles become single-peaked, with increasingly sloping wings and higher kurtosis, as they become more dispersion supported.

In Figure~\ref{fig:data_schematic}, we show the unresolved HI profiles of two galaxies in our observed sample, both with $M_{\rm star}\sim 10^8 M_{\odot}$. One of these galaxies exhibits a double-horned profile with low kurtosis, (qualitatively similar to \texttt{m11b}), while the other produces a single-peaked profile with higher kurtosis (qualitatively similar to \texttt{m11a}). Spatially resolved HI kinematics are not available for our observed sample, so we instead show the galaxies' optical images from the NASA-Sloan Atlas, a reprocessing of SDSS DR8 \citep{Blanton_2011}. The optical morphology of the target with the double-peaked 21-cm profile is clearly disky, while that of the target with the Gaussian-like profile appears clumpy and irregular. Stellar morphology from optical images is an imperfect proxy for HI morphology, and we do not carry out any quantitative analysis on optical images. However, we find that on average, galaxies with double-peaked HI profiles exhibit diskier optical morphologies upon visual inspection than those with single-peaked profiles.

\begin{figure}
\includegraphics[width=\columnwidth]{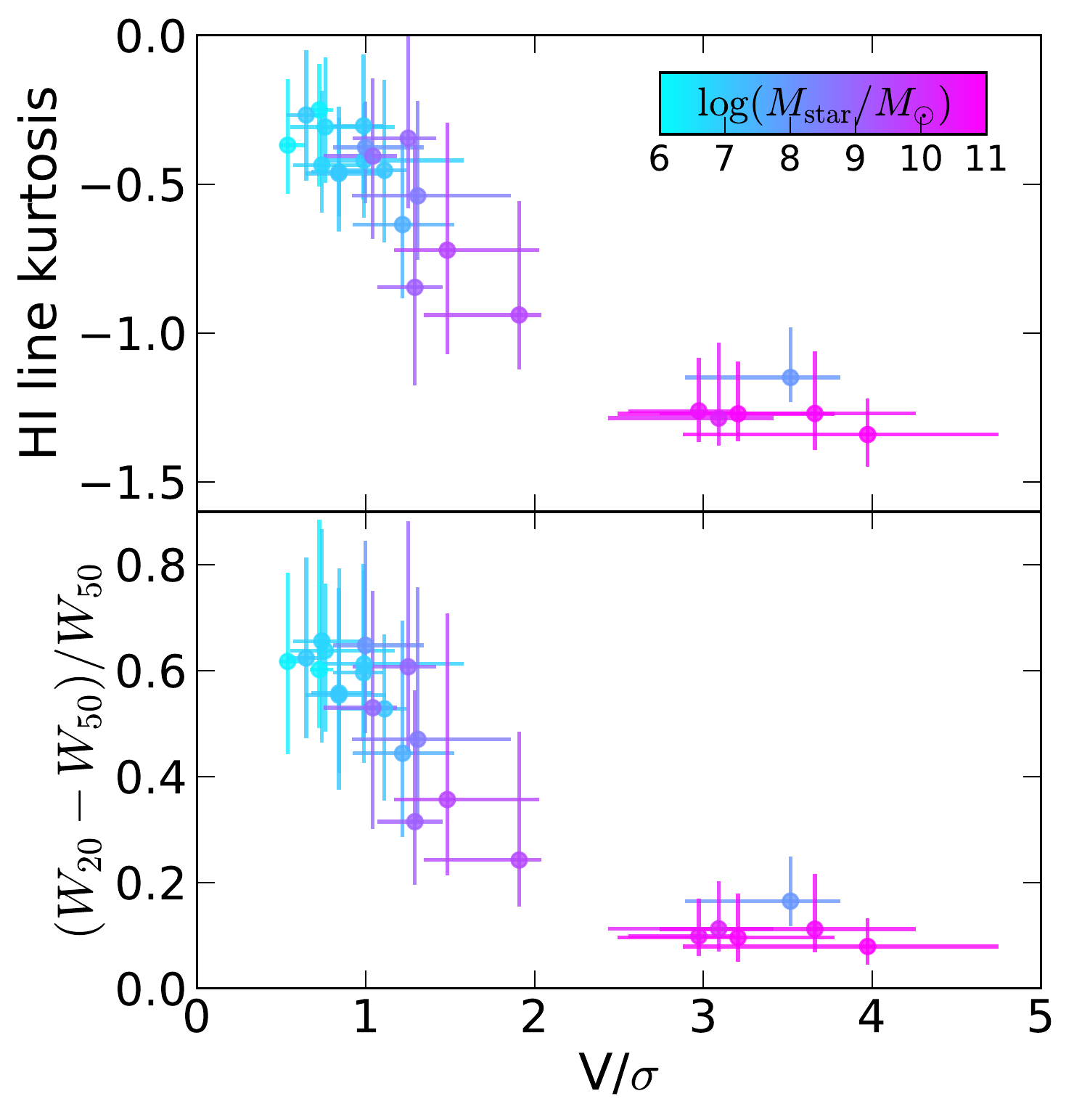}
\caption{Kurtosis (top) and steepness of the wings (bottom) of unresolved 21-cm line profiles of simulated galaxies vs. $V/\sigma$ (from mock-IFU measurements; Equation~\ref{eqn:v_sig}). Error bars show 68\% scatter across 100 different viewing angles. Disky, rotationally supported galaxies with higher $V/\sigma$ have systematically lower kurtosis ($\kappa \lesssim -1$) and steeper profile wings ($(W_{20}-W_{50})/W_{50}\lesssim 0.2$) than dispersion supported galaxies. Rotation- and dispersion-supported galaxies can thus be separated statistically based only on their unresolved line profile shape.}
\label{fig:vsig}
\end{figure}

To determine quantitatively how the shape of the unresolved HI line is related to a galaxy's \textit{resolved} gas kinematics, we mock-observe all our simulations from 100 random viewing angles and calculate the kurtosis, $\kappa$, wing steepness, $\Delta W$, and the degree of rotation vs. dispersion support, $V/\sigma$, for each viewing angle. To calculate $V/\sigma$ for a given viewing angle, we first construct resolved mock HI velocity maps with spatial resolution 0.1 kpc (see \citetalias{ElBadry_2018} for details) and then define \begin{equation}
\label{eqn:v_sig}
V/\sigma=\frac{\sum_{j}m_{j}\left|V_{{\rm los},j}\right|}{\sum_{j}m_{j}\sigma_{{\rm los},j}}.
\end{equation}
Here $m_{j}$ is the HI mass in a given spatial pixel, $V_{{\rm los},j}$ and $\sigma_{{\rm los},j}$ are the HI mass-weighted mean line-of-sight velocity and velocity dispersion in a pixel, and the sum is over all spatial pixels in the data cube. We take this definition from \citet{Obreja_2016}. Although $V/\sigma$ as calculated in Equation~\ref{eqn:v_sig} is dependent on the size of spatial pixels; i.e., the beam size in resolved observations, it has the advantage of being defined in terms of quantities measured directly from the simulation, without requiring any kinematic model fitting. Like $\kappa$ and $\Delta W$, $V/\sigma$ is a line-of-sight dependent quantity. Projection effects can cause rotational velocities to contribute significantly to $\sigma$, so $V/\sigma$ values calculated from Equation~\ref{eqn:v_sig} are typically lower than values calculated from cylindrically-averaged 3D quantities \citepalias[see][]{ElBadry_2018}. 

In Figure~\ref{fig:vsig}, we plot $\kappa$ and $\Delta W$ for our simulated galaxies as a function of $V/\sigma$. For all quantities, we show the median and middle 68\% values across 100 random viewing angles. $\kappa$ and $\Delta W$ are measured from mock unresolved HI profiles with a SNR of 25, the median value for our observed sample. In agreement with the qualitative trend apparent in Figure~\ref{fig:schematic}, we find that $\kappa$ and $\Delta W$ decrease nearly monotonically with increasing $V/\sigma$, indicating that these quantities can be used as a proxy for $V/\sigma$. We find that $\kappa$ and $\Delta W$ are also both anticorrelated with the intrinsic, three-dimensional $V_{\rm rot}/\sigma$, as defined in \citetalias{ElBadry_2018}. However, the correlation is somewhat weaker than the correlation with $V/\sigma$ as defined in Equation~\ref{eqn:v_sig} because the unambiguously rotation-supported galaxies all have $\kappa \sim -1.4$ and $\Delta W \sim 0.1$ but have a wider range of $V_{\rm rot}/\sigma$, with $6 \lesssim V_{\rm rot}/\sigma \lesssim 12$.

Figure~\ref{fig:vsig} also shows that, for the simulated galaxies, $V/\sigma$, $\kappa$, and $\Delta W$ are all strong functions of mass, with higher-mass galaxies having higher $V/\sigma$ and lower $\kappa$ and $\Delta W$.  However, the trend of decreasing kurtosis and increasingly steep profile wings at higher $V/\sigma$ is \textit{not} simply a result of the joint dependence on mass: the lowest-mass galaxy in our sample that is strongly rotation-supported (\texttt{m11b}, with $M_{\rm star}=10^8 M_{\odot}$) still has low $\kappa$ and $\Delta W$. For more detailed discussion of trends in gas kinematics with mass for this sample of simulated galaxies, we refer to \citetalias{ElBadry_2018}. 

\begin{figure}	
\includegraphics[width=\columnwidth]{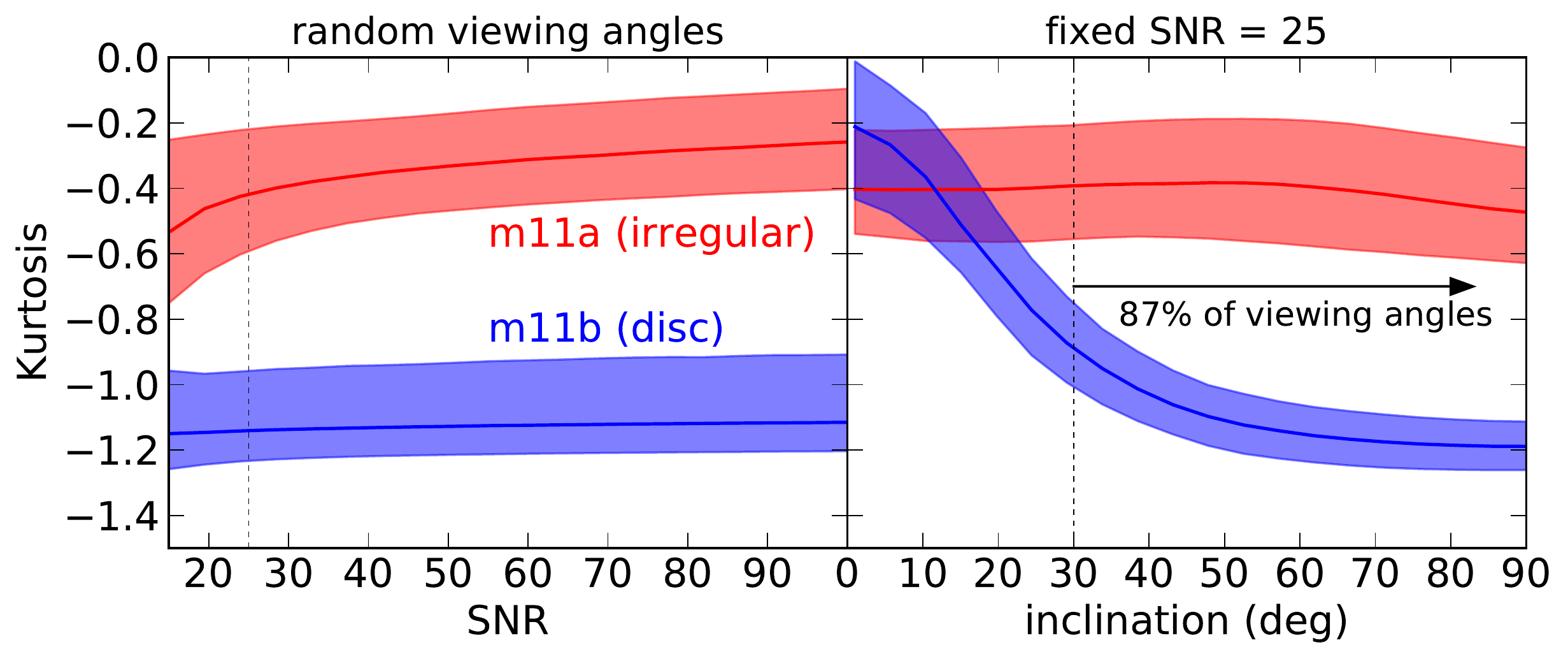}
    \caption{\textbf{Left}: Effect of signal-to-noise ratio on 21-cm line profile kurtosis measurements for a simulated disky, rotationally supported galaxy (blue) and an irregular, dispersion supported galaxy (red). Dashed line shows $\rm SNR=25$, the median value for our observed sample. At each SNR, we mock-observe each galaxy from 100 random viewing angles and generate 100 different noise realizations for each viewing angle. We plot the median and middle 68\% scatter for the resulting $10^4$ profiles generated at each SNR. \textbf{Right}: Here we vary inclination while fixing $\rm SNR=25$, the median for the observed sample in Figures~\ref{fig:kurt_vs_mstar} and~\ref{fig:W20_W50}. $\kappa$ is robust to changes in SNR down to ${\rm SNR}\sim 15$. $\kappa$ does vary with inclination for disky systems, but the effects are large only for rare, nearly face-on viewing angles.}
\label{fig:snr}
\end{figure}

Although the non-negligible 68\% scatter in Figure~\ref{fig:vsig} shows that $\kappa$ and $\Delta W$ do vary across viewing angles, rotation- and dispersion-supported systems are still separated on average. In Figure~\ref{fig:snr}, we show explicitly how $\kappa$ varies with SNR (left) and inclination (right). For each inclination and SNR value, we compute the kurtosis of $10^4$ mock HI profiles for \texttt{m11a} and \texttt{m11b}, two of the galaxies shown in Figure~\ref{fig:schematic}. These galaxies both have $M_{\rm star} \approx 10^8 M_{\odot}$ and $M_{\rm 200m}\approx 10^{10.7}M_{\odot}$, but \texttt{m11b} is rotation-supported while \texttt{m11a} is dispersion-supported. The left panel of Figure~\ref{fig:snr} shows that, at ${\rm SNR} > 15$ (the minimum SNR of our observed sample), kurtosis measurements for a given system are robust: the typical $\kappa$ values for the two galaxies remain well-separated, and the spread of $\kappa$ values across different SNR is less than that across different viewing angles at fixed SNR. 

The right panel of Figure~\ref{fig:snr} shows how kurtosis varies with inclination angle, which is defined in terms of the net angular momentum vector of HI. For the disky, rotation-supported galaxy, kurtosis varies significantly with inclination: it has $\kappa \sim -1.2$ when viewed edge on, but $\kappa \sim 0$ when viewed face on. This is unsurprising, since the face-on disk has no rotation projected along the line of sight; generically, rotation-supported disk galaxies will produce single-peaked, Gaussian-like profiles if viewed completely face-on.

Fortunately, most rotation-supported galaxies will nevertheless produce 21-cm profiles with low kurtosis, because face-on viewing angles are intrinsically rare: for random disk orientations, the probability distribution of inclinations is $p\left(i\right)\,{\rm d}i=\sin i\,{\rm d}i$, meaning that far more galaxies are observed nearly edge-on than nearly face-on. Because inclination effects on kurtosis measurements only become significant at $i \lesssim 40$ degrees (representing 23\% of viewing angles), rotationally supported galaxies will produce unresolved HI profiles with low kurtosis for most viewing angles.

For the dispersion-supported galaxy, there is no significant trend in $\kappa$ with inclination, though there is still significant scatter across different viewing angles at fixed inclination. This is unsurprising, since dispersion-supported systems have no unambiguously preferred kinematic axis. Most lines of sight for such systems produce single-peaked line profiles roughly resembling Gaussians, but large-scale inflows and outflows of HI, as well as irregular HI distributions, can produce a variety of irregular and asymmetric profiles. 

Although it is not shown in Figure~\ref{fig:snr}, we find that $\Delta W$ has a similar scaling with SNR and inclination to $\kappa$. Due to viewing angle effects, the mapping from line shape as quantified by $\kappa$ or $\Delta W$ to $V/\sigma$ is not unique. However, since real galaxies should have random orientations, the observed distribution of $\kappa$ and $\Delta W$ values places a statistical constraint on the observed population's distribution of $V/\sigma$ values, even though it is not always possible to unambiguously measure the kinematics of a particular galaxy from its unresolved line profile.

\subsubsection{Effect of rotation curve shape}
The detailed shape of a galaxy's unresolved 21-cm line profile depends not only on rotation vs. dispersion support, but also on the shape of its rotation curve. We investigate this dependence in detail in Appendix~\ref{sec:toy_model}. Our primary conclusion is that while a double-peaked 21-cm profile is almost always indicative of rotational support, a single-peaked, Gaussian-like profile can arise either due to dispersion support \textit{or} due to a linearly rising rotation curve: if a galaxy's circular velocity curve rises sufficiently slowly and does not begin to flatten in the region occupied by the galaxy, a single-peaked profile can be produced even in the limit where HI is completely rotationally supported and follows the circular velocity curve, $v_c(r)=\sqrt{r\times{\rm d\Phi}/{\rm d}r}$. For the simulated galaxies, we can measure the HI rotation curve and $v_c(r)$ directly, and we find that Gaussian-like profiles are due to dispersion support, not linearly rising rotation curves. However, for observed galaxies with Gaussian-like HI profiles, it is generally not possible to distinguish between a dispersion-supported system and a rotation-supported systems with a linearly rising rotation curve.

\section{Comparison to observations}
\label{sec:obs_comp}
\begin{figure}
\includegraphics[width=\columnwidth]{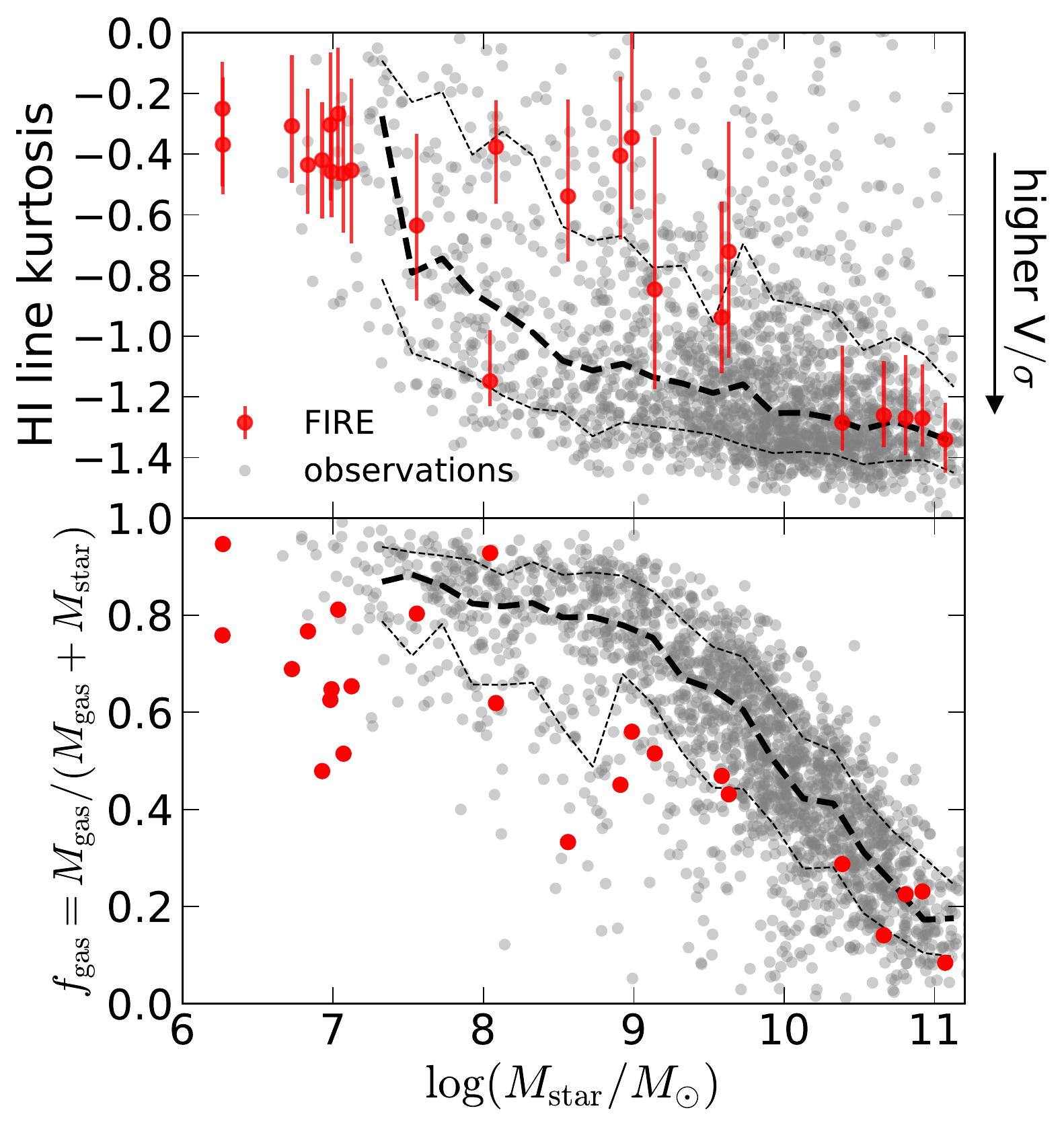}
\caption{\textbf{Top}: 21-cm line kurtosis measurements for simulated galaxies compared to $\sim$2000 observed galaxies from \citetalias{Bradford_2015}. Error bars show 68\% scatter across 100 random viewing angles; dashed lines show median and 68\% scatter for observed galaxies in each mass bin. At $M_{\rm star} > 10^{10} M_{\odot}$, where all the simulated galaxies have rotation-supported HI disks, the simulated galaxies produce line profiles with kurtosis values very similar to the observed galaxy population. At $M_{\rm star} = 10^{8-10} M_{\odot}$, most of the simulated profiles have higher $\kappa$ than the observed population, corresponding to lower $V/\sigma$ (see Figure~\ref{fig:vsig}).
\textbf{Bottom}: Gas fractions. Most low-mass simulated galaxies with higher $\kappa$ than the median of the observed sample also have somewhat lower gas fractions than typical observed galaxies. }
\label{fig:kurt_vs_mstar}
\end{figure}

\subsection{Kurtosis}
We now compare the unresolved HI profiles generated from our simulations to those of real galaxies. Our observed sample consists of 2002 21-cm HI profiles with ${\rm SNR} > 15$. Almost all of these profiles were presented in \citetalias{Bradford_2015}; we also include in our sample observations of few dozen  low-mass galaxies that were obtained since the publication of that work and were reduced following the same procedure described in \citetalias{Bradford_2015}. The majority of this observational sample is completely morphologically blind. 210 of the profiles first presented by \citetalias{Bradford_2015} were selected with a weak selection criterion against galaxies with inclinations near 0. However, we find the full observational sample has a distribution of inclinations (computed from optical axis ratios by \citetalias{Bradford_2015}) consistent with random orientations.  

We compute the kurtosis of the smoothed emission regions of these profiles, as defined in Section~\ref{sec:mock_obs}, using Equation~\ref{eqn:kurt}. Stellar masses were taken from the NASA-Sloan atlas \citep{Blanton_2011}. The median SNR of these data is 25, so we adopt ${\rm SNR=25}$ in generating mock HI profiles for the simulated galaxies. In this observational sample, there is no significant correlation between SNR and $\kappa$ or $M_{\rm star}$. 

In Figure~\ref{fig:kurt_vs_mstar}, we show the distribution of kurtosis values for these galaxies as a function of mass, overplotting measurements for our simulated galaxies. At $M_{\rm star} \gtrsim 10^{10} M_{\odot}$, most observed galaxies have $\kappa \lesssim -1$, indicating that they have rotation-supported kinematics. This is also true for the simulated galaxies, which fall on top of the median observed relation. At this mass scale, the scatter in the kurtosis measurements of different profiles for individual galaxies, which is due primarily to different viewing angles, is almost as large as all the scatter in the observed sample. This suggests that gas-rich galaxies at $M_{\rm star} \gtrsim 10^{10} M_{\odot}$ likely all have similar rotation-supported HI kinematics.

At lower masses, particularly $M_{\rm star} = 10^{8-10} M_{\odot}$, most of the simulated galaxies have significantly higher kurtosis than at Milky Way (MW) masses, with $\kappa \sim -0.5$. A similar trend toward higher $\kappa$ at lower masses is evident in the observations, but the trend is less pronounced than for the simulated galaxies: more than half of the observed galaxies have $\kappa < -1$ over this mass range, indicating rotation-supported kinematics. Of the 8 simulated galaxies with $M_{\rm star} = 10^{8-10} M_{\odot}$, 7 have median $\kappa$ values above the median observed $\kappa - M_{\rm star}$ relation. The only exception is \texttt{m11b}, which has a rotation-supported disk (Figure~\ref{fig:schematic}) and falls below the median observed relation. 

\subsection{Gas Fractions}
In the bottom panel of Figure~\ref{fig:kurt_vs_mstar}, we compare the gas fractions of the simulated galaxies to observations. For both the observed and simulated galaxies, we define $M_{{\rm gas}} \equiv 1.4M_{{\rm HI}}$, where the factor of 1.4 accounts for helium, and then plot $f_{{\rm gas}}=M_{{\rm gas}}/(M_{{\rm gas}}+M_{{\rm star}})$. For both the observed and simulated samples, $f_{\rm gas}$ decreases with increasing $M_{\rm star}$. The simulations with $M_{\rm star} > 10^{10} M_{\odot}$ have gas fractions similar to the median of the observed population, but at lower masses, most of the simulated galaxies have somewhat lower $f_{\rm gas}$ than the median observed relation. 

In particular, most of the low-mass simulated galaxies with higher-than-average $\kappa$ (i.e., enhanced dispersion support) also have lower gas fractions than the median of the observed population. The lowest-mass simulated galaxy with a clear rotation-supported disk (\texttt{m11b}, with $M_{\rm star}=10^{8} M_{\odot}$) also has a higher-than-average gas fraction ($f_{\rm gas} = 0.94$). This suggests that for the simulated galaxies, increased dispersion supported is likely connected to reduced gas content. Intriguingly, we find no correlation between $f_{\rm gas}$ and $\kappa$ at fixed mass in the observed sample: at all masses, observed galaxies with lower-than-average gas fractions appear to have similar gas kinematics to those with higher-than-average gas fractions. We return to this issue in Section~\ref{sec:biases}.

\begin{figure}	
\includegraphics[width=\columnwidth]{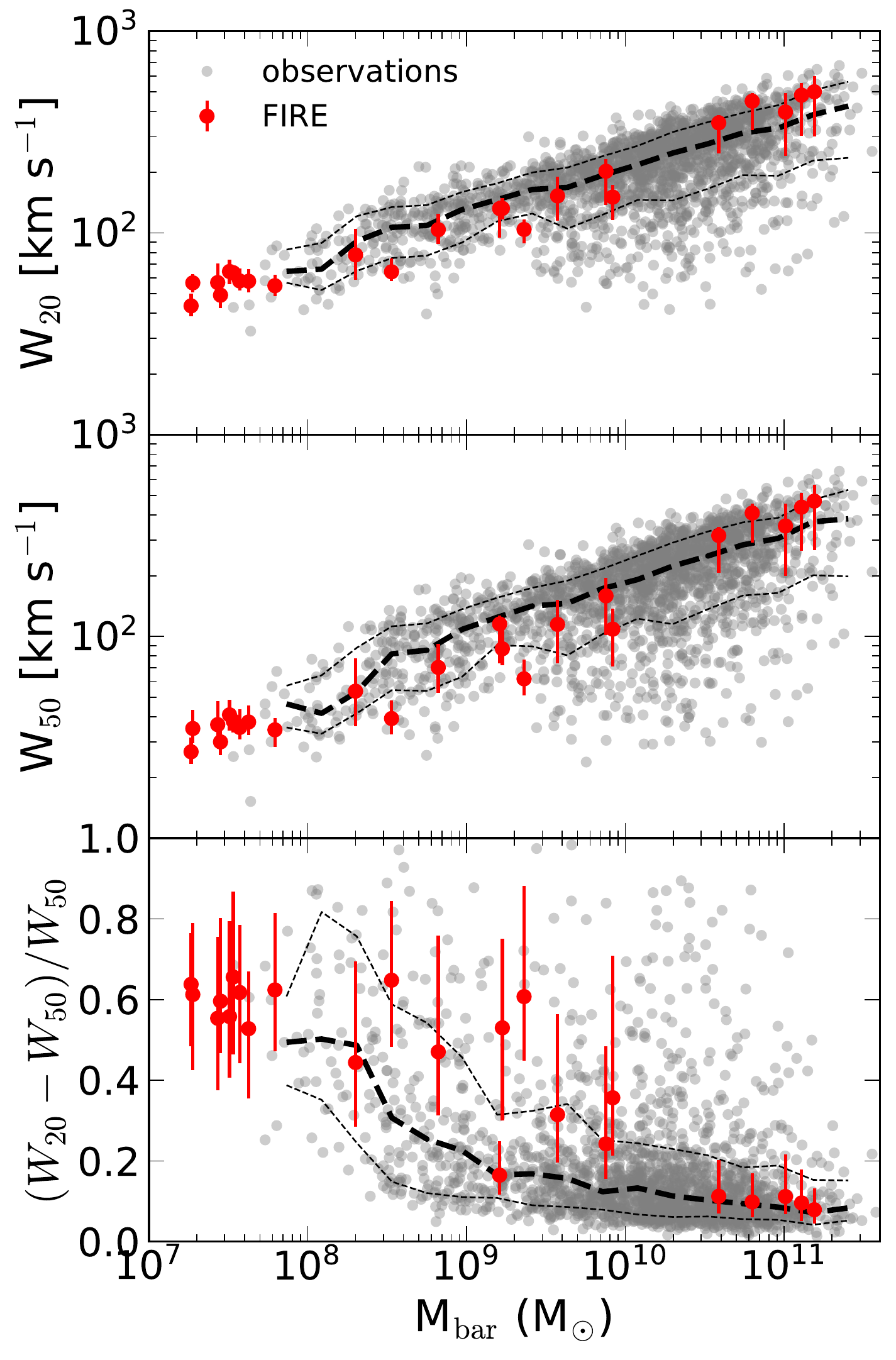}
    \caption{Spatially-unresolved baryonic Tully-Fisher relation as probed by $W_{20}$ (top) and $W_{50}$ (middle), without inclination corrections. Error bars show 68\% scatter across 100 viewing angles. Bottom panel shows $\Delta W = (W_{20}-W_{50})/W_{50}$, an estimate of the slope of the wings of the HI profile. The simulations reproduce the observed BTFR as probed by $W_{20}$ well. At $\sim$MW masses ($M_{\rm bar} > 10^{10.5}M_{\odot}$), the simulated galaxies also have similar $W_{50}$ and $\Delta W$ to the observed galaxies. At lower masses, most simulated galaxies have $W_{20}$ similar to the observed population but lower $W_{50}$ and thus higher $\Delta W$, indicating less-steep profile wings and higher dispersion than average for the observed galaxies.}
\label{fig:W20_W50}
\end{figure}

\subsection{Line Widths and BTFR}
The width of a galaxy's unresolved 21-cm line, as quantified by $W_{20}$ or $W_{50}$, is typically used as a diagnostic of its maximum circular velocity, and thus, the depth of its gravitational potential. Velocity widths are used to construct the BTFR for gas-rich galaxies \citep[e.g.][]{McGaugh_2000}. In this context, individual line widths are to first order agnostic to rotation vs. dispersion support, since the range of line-of-sight velocities in a virialized system depends primarily on the depth of the gravitational potential. Different BTFR slopes can be constructed depending on whether $W_{20}$ or $W_{50}$ is used; see \citet{Bradford_2016} for a review.

To compare the BTFR for our simulated galaxies with observations, we follow \citetalias{Bradford_2015} in defining baryonic masses for the simulated galaxies as $M_{\rm bar} = M_{\rm star}+1.4M_{\rm HI}$. We correct the observed $W_{20}$ and $W_{50}$ values for redshift broadening by dividing them by $(1+z)$; this correction is always small because $z\ll 1$. Most previous works on the BTFR have also attempted to correct for inclination effects by dividing observed line widths by $\sin i$, where inclination is calculated from galaxies' projected optical axis ratios with the assumption of an intrinsic thin-disk morphology. Such corrections can be problematic for low-mass galaxies, both because optical axis ratios are an imperfect tracer of HI morphology, and because there is no guarantee that galaxies' HI morphologies are intrinsically disky. We therefore do not apply any inclination correction to the observed or simulated line widths, instead accounting for the random orientations of the observed galaxies by mock-observing the simulated galaxies along many random lines of sight. As a result, the intrinsic scatter in the observed line width -- mass relation is larger than that found when an inclination correction is applied (see \citetalias{Bradford_2015}).

In Figure~\ref{fig:W20_W50}, we compare the BTFR predicted by our simulations as quantified by $W_{20}$ (top) and $W_{50}$ (middle) to that of the observational sample from \citetalias{Bradford_2015}. At $M_{\rm bar} > 10^{10.5} M_{\odot}$, the simulated HI profiles have both $W_{\rm 20}$ and $W_{50}$ in good agreement with the observed sample. At lower masses, the simulated $W_{20}$ values remain in reasonable agreement with the observed values, while the simulated $W_{50}$ values are on average lower than the median of the observed values. Because our simulated galaxies -- and the observed galaxies, to a lesser extent -- preferentially produce 21-cm profiles with sloping wings at lower masses, the slope of the BTFR measured by $W_{50}$ is steeper than that measured by $W_{20}$.

The bottom panel of Figure~\ref{fig:W20_W50} shows $\Delta W$. At $M_{\rm bar} \gtrsim 10^{10.5} M_{\odot}$, values for the simulated and observed galaxies agree well. At $M_{\rm bar}=10^{8.5-10} M_{\odot}$, most of the simulated galaxies have higher $\Delta W$, corresponding to shallower HI profile wings, than typical observed galaxies. Consistent with Figure~\ref{fig:kurt_vs_mstar}, the higher $\Delta W$ for low-mass simulated galaxies suggests that they have enhanced dispersion support compared to the observed galaxies. 

Measured from unresolved line widths, the BTFR is relatively insensitive to rotation vs. dispersion support. This is in contrast to measurements from spatially resolved gas kinematics \citep[e.g.][]{Simons_2015, Bloom_2017}, which can distinguish between rotation and dispersion. For unresolved observations, simulated galaxies can be expected to match the observed BTFR as long as they fall on a realistic $M_{\rm bar}-M_{\rm halo}$ relation and are approximately virialized, irrespective of the mode of kinematic support. The simulated galaxies reproduce the observed BTFR as measured by $W_{20}$ well, despite the apparent excess dispersion support in low-mass galaxies. On the other hand, the dispersion-supported simulated galaxies fall slightly below the observed BTFR as measured by $W_{50}$. This occurs because $W_{50}$ is significantly less than $W_{20}$ for a dispersion-supported system with a Gaussian LOSVD, but the two quantities are similar for a rotation-supported systems with steep wings. This suggests that the scatter in the BTFR can be reduced by measuring $W_{20}$ as opposed to $W_{50}$, or by inflating $W_{50}$ values for targets with large $\Delta W$ and $\kappa$. Previous studies with observational data \citep{Bradford_2016, Papastergis_2016} have indeed found this to be the case. Incorporating measurements of line shape as quantified by $\kappa$ and $\Delta W$ in the observed BTFR will likely prove useful in minimizing the relation's scatter for the purpose of using it as a redshift-independent distance estimator \citep[e.g.][]{Obreschkow_2013}.

\section{Summary and Discussion}
\label{sec:conclusion}

We have investigated the connection between the shape of galaxies' spatially unresolved 21-cm line profiles and the kinematics of their gas. Rotation-supported galaxies typically produce double-peaked line profiles with steep wings, while dispersion-supported galaxies produce single-peaked Gaussian-like profiles with more sloping wings (see Figure~\ref{fig:schematic}). Using a suite of simulations from the FIRE project, we showed that the kurtosis and wing steepness of galaxies' unresolved HI lines can thus be used as a proxy for $V/\sigma$ (Figure~\ref{fig:vsig}). Although the mapping between line shape and $V/\sigma$ also depends on viewing angle (Figure~\ref{fig:snr}) and on the shape of galaxies' rotation curves (Appendix~\ref{sec:toy_model}), marginalization over these parameters with a large observational sample provides statistical constraints on the degree of rotational support in the local galaxy population, without requiring spatially resolved data. 

Unresolved 21-cm line profiles provide a homogeneous, unbiased observational sample with which to test the predictions of simulations. Because mock observations of unresolved HI are straightforward to carry out with minimal assumptions and no model fitting, they allow for robust comparison of simulations and observational data on equal terms.  We compared mock 21-cm line profiles of our simulated galaxies to those of a sample of $\sim$2000 low-redshift observed galaxies, quantifying their shape through the measured kurtosis (Figure~\ref{fig:kurt_vs_mstar}) and the difference between their measured 20\% and 50\% velocity widths (Figure~\ref{fig:W20_W50}). The primary results of these comparisons are as follows: 
\begin{enumerate}
\item \textit{Kinematics in Milky Way-mass galaxies}: At $M_{\rm star} = 10^{10-11}M_{\odot}$, the simulated galaxies all have rotation-supported gas disks and produce HI profiles with shapes in excellent agreement with the observed population. At this mass scale, the scatter in observed kurtosis values is small enough to be explained almost entirely by inclination effects, indicating that most HI-rich galaxies with $M_{\rm star} = 10^{10-11} M_{\odot}$ have similar, rotation-supported gas kinematics.

\item \textit{Kinematics in lower-mass galaxies}: At $M_{\rm star}=10^{8-10} M_{\odot}$, most of the simulated galaxies have higher kurtosis values and less-steep wings than more massive galaxies. This is true for the observed sample as well. However, the simulated lower-mass galaxies appear to have somewhat more dispersion support than the observed systems (Figures~\ref{fig:kurt_vs_mstar} and ~\ref{fig:W20_W50}). This is not universally true: one low-mass galaxy (\texttt{m11b}, with $M_{\rm star} = 10^{8} M_{\odot}$) does have a kinematically cold, rotationally supported gas disk and produces a characteristic double-horned HI profile. However, 7 of the 8 simulated galaxies in our sample with $M_{\rm star}=10^{8-10}M_{\odot}$ produce HI profiles with higher kurtosis and shallower wings than the median of the observed sample.\footnote{We also note that the \texttt{m11b} halo has a relatively high spin parameter, with $\lambda = 0.077$, which is $2\sigma$ above the mean for all dark matter halos \citep{Bullock_2001}; this may be related to its diskier-than-average morphology and kinematics.}
\item \textit{Gas fractions}: The simulated MW-way mass galaxies have gas fractions similar to the observed galaxies. At lower masses ($M_{\rm star}=10^{8-10}M_{\odot}$), the gas fractions of most simulated galaxies are somewhat lower than the median of the observed population (Figure~\ref{fig:kurt_vs_mstar}). For the simulated low-mass galaxies, $f_{\rm gas}$ and $V/\sigma$ are correlated at fixed mass: galaxies with lower-than-average gas fractions have higher-than-average $\kappa$ and $\Delta W$, indicating enhanced dispersion support. We do not find such a correlation for the observed galaxies.
\item \textit{Baryonic Tully-Fisher relation}: At all masses, the simulated galaxies have 20\% line-widths in good agreement with the observed baryonic Tully-Fisher relation, irrespective of whether they are supported by rotation or dispersion (Figure~\ref{fig:W20_W50}). Global line widths are primarily probes of galaxies' virial masses and are agnostic to the mode of kinematic support.  
\end{enumerate}

At the lowest mass scales probed by our simulations ($M_{\rm star} \simeq 10^{6-7.5} M_{\odot}$), the observed sample does not contain enough observed galaxies for a robust comparison between the simulations and observations. Like the simulated galaxies, the lowest-mass observed galaxies preferentially have single-peaked profiles with high $\kappa$ and $\Delta W$, but this could be explained by either increased dispersion support or linearly rising circular velocity curves (see Appendix~\ref{sec:toy_model}).

\subsection{Possible selection effects}
\label{sec:biases}
We note that although no explicit morphological selection criteria were used in constructing our observed sample, some selection effects could be present. In particular, our observed galaxies are all detected in both HI and optical light, which could select against (a) low surface brightness or (b) HI-poor galaxies. Many of our low-mass galaxies do have low surface brightness \citep{Chan_2017}. We do not, however, expect a selection effect against low surface brightness galaxies in our observed sample to significantly bias our observationally inferred kinematics, as observed low surface brightness galaxies typically have ordinary levels of rotational support in their gas kinematics \citep{deBlok_2001, Oh_2015}. 

Because most of the dispersion-supported galaxies in our simulation sample also have somewhat lower gas fractions than typical observed galaxies with similar mass, one might expect selection effects against HI-poor galaxies to introduce biases against dispersion-supported systems in the observed sample. However, the lack of any correlation between gas fraction and kinematics in the observed sample suggests that the excess dispersion found in the simulated galaxies is \textit{not} primarily due to poor observational completeness at low $f_{\rm gas}$: if it were, observed galaxies with lower-than-average $f_{\rm gas}$ would show enhanced dispersion support. 

We also consider the possibility of biases in the sample of simulated galaxies. The primary selection effect in our simulation sample is that we target halos that are isolated at $z=0$, meaning that they have no more massive neighbors within at least $3\times R_{\rm 200m}$. We have verified that the sample spans a normal range of  halo spin, concentration and formation time \citepalias{ElBadry_2018}, and previous work has also found that targeting isolated halos does not introduce significant biases in most halo properties \citep{Onorbe_2014}. 

To test explicitly whether our selection of isolated halos introduces biases in the galaxy properties studied here, we repeated our analysis with only the 923 galaxies in our observational sample (out of 2002 total) that were found to be isolated by \citetalias{Bradford_2015}. We found this isolated subsample to have similar $f_{\rm gas}$, $\kappa$, and $\Delta W$ to the full observational sample.\footnote{This is true only because the observed galaxies all retain enough cold gas to be detected in HI. Stripped galaxies in dense cluster environments likely do have different gas properties than isolated galaxies, but such targets do no enter our observational sample.} We therefore conclude that simulating halos that are isolated at late times does not introduce significant biases for the galaxy properties studied here. 

\subsection{Excess dispersion in low-mass galaxies}

The \textit{resolved} gas properties of the simulated galaxies studied here were previously investigated and compared to observations by \citetalias{ElBadry_2018}. In qualitative agreement with our results, that work found most galaxies with $M_{\rm star}=10^{8-10}M_{\odot}$ to be somewhat less rotationally supported than typical observed galaxies at the same mass scale (e.g., their Figure 15). Due to imperfectly understood selection effects in the observed samples to which they compared, \citetalias{ElBadry_2018} concluded that there were hints of enhanced dispersion support in the simulations at low masses, but they could not verify the disagreement to be robust. Our results also find evidence of excess dispersion support at low masses, but with comparison to a larger, morphologically unbiased observational sample. We emphasize that although our simulated low-mass galaxies are \textit{on average} more dispersion-supported than the observed population, they still fall in regions of parameter space populated with real galaxies; i.e., the simulations do not produce galaxies in observationally ``forbidden'' regions of parameter space.

The mass scale at which we find the most enhanced gas dispersion in the simulated galaxies compared to observations is precisely the scale at which feedback-driven outflows have been shown to most strongly alter galaxies' gravitational potentials \citep{DiCintio_2014, Chan_2015, ElBadry_2016, Bullock_2017}. In the FIRE simulations, dynamical heating through a feedback-driven, time-variable potential at this mass scale has been shown to produce dark matter cores \citep{Chan_2015}, low central stellar surface densities \citep{Chan_2017}, and dispersion-supported stellar dynamics with primarily radial orbits \citep{ElBadry_2017}. 

The same large-scale outflows that lead to core formation (a) continually inject energy into low-mass galaxies' ISM, making it difficult for gas to settle into a rotationally supported disk \citepalias{ElBadry_2018} and (b) produce very bursty star formation histories at low masses. This suggests that the excess gas dispersion we find at this mass scale is likely connected to tensions between the simulations and observations identified by previous work. \citet{Sparre_2017} showed that FIRE galaxies with $M_{\rm star}\lesssim 10^{9.5} M_{\odot}$ likely have star formation histories that at $z=0$ are on average burstier than those of typical low-mass field galaxies (though observations also imply nontrivial SFH burstiness at this mass scale; see \citealt{Weisz_2012}), and \citet{Chan_2017} showed that galaxies from the FIRE simulations with $M_{\rm star}=10^{7-9} M_{\odot}$ have, on average, lower surface brightnesses than typical observed galaxies at the same stellar mass. 

One might therefore expect that low-mass  galaxies with rotation-supported gas kinematics would have calmer star formation histories and smaller cores than similar-mass galaxies with dispersion-supported gas; we investigate this further in Appendix~\ref{sec:sfhs}. We find that the lowest-mass galaxy in our simulation sample with a clear rotation-supported disk has a star formation history (SFH) that is substantially less bursty at late times than the SFHs of the dispersion-supported simulated galaxies at the same mass scale. It nevertheless has a dark matter core at $z=0$, likely due to burstier star formation at early times, but the core is smaller than in dispersion-supported systems at the same mass scale. 

The apparent excess of dispersion support in the low-mass simulated galaxies can serve as a useful constraint on feedback models and may motivate modifications to the FIRE model or the incorporation of additional physical processes in the future. We first consider the possibility that the FIRE model produces feedback that is in some sense ``too strong'' at low masses. Because feedback energetics are taken directly from stellar evolution calculations, there is little room to arbitrarily reduce or increase the overall feedback energy budget.\footnote{In \citet{Hopkins_2017} and Hopkins et al. (in prep) we have nevertheless carried out tests of the effects of modified feedback recipes on the burstiness of star formation in dwarfs. While reducing the feedback energy budget can make galaxies somewhat less bursty and morphologically diskier at fixed halo mass, doing so increases galaxy stellar masses, such that at fixed stellar mass, burstiness is not significantly changed. Significantly changing the feedback energy budget also produces a $M_{\rm star} - M_{\rm halo}$ relation in worse agreement with constraints from abundance matching than the default FIRE model.} It is possible, however, that changing details in the implementation of physical processes that are currently modeled approximately (e.g., adding on-the-fly radiative transport or more explicit treatment of HII regions) could change the burstiness of star formation. For example, if feedback can create hot ``chimneys'' surrounding star forming regions such that localized outflows can vent energy out of the disk \citep[e.g.][]{Ceverino_2009, Hopkins_2012}, the effects of feedback on the bulk of a galaxy's gas may be less catastrophic than when energy and momentum are primarily deposited in cold gas. Alternatively, if radiative feedback can suppress star formation locally on timescales shorter than the free-fall time in bound gas clouds \citep[e.g.][]{FG_2018}, star formation can become less spatially and temporally clustered, leading to weaker outflows and less dispersion-supported gas.

Simply turning off some feedback sources (e.g. radiative feedback or stellar winds) does not seem to reduce SFH burstiness, as it allows more stars to form before collapse is regulated by supernovae \citep{Hopkins_2017}. As long as star formation is restricted to self-gravitating gas clouds, varying the local star formation efficiency or the critical density above which star formation is allowed to occur also does not significantly change galaxy SFHs in the default FIRE model, since star formation proceeds until feedback produces enough energy to regulate it \citep{Hopkins_2017}. It may thus if fact be easier to facilitate disk formation by incorporating additional feedback processes, albeit processes that act less violently than supernovae, than by reducing the types of feedback sources or their energy budget. For example, \citet{Chen_2016} found that including cosmic ray diffusion in galaxy formation simulations caused their low-mass galaxies to transition from being dispersion-supported, with bursty SFHs, to rotation-supported, with smoother SFHs. Further investigation of the effects of changes in the details of feedback implementations on SFH burstiness and galaxy morphology represents a promising avenue for future work. We note, for example, that feedback resulting in weaker large-scale outflows in low-mass galaxies could also alleviate the difficulty identified by \citet{AnglesAlcazar_2017} of growing massive black holes in the FIRE simulations at early times.

The fact that our simulations do produce some low-mass galaxies with rotation-supported disks also suggests that at fixed mass, morphology is driven by a galaxy's formation history and/or environment. This warrants further investigation of the assembly histories of our low-mass galaxies and the primary factors that drive them to become rotation- or dispersion-supported, as well as simulation of low-mass galaxies in a wider range of environments. 

\section*{Acknowledgements}
We thank the referee, Danail Obreschkow, for a constructive report and Alyson Brooks and Pascal Elahi for helpful discussions. 
KE acknowledges support from a Berkeley graduate fellowship, a Hellman award for graduate study, and an NSF Graduate Research Fellowship. 
EQ and KE are supported by a Simons Investigator Award from the Simons Foundation and by NSF grant AST-1715070.
MBK acknowledges support from NSF grant AST-1517226 and from NASA grants NNX17AG29G and HST-AR-13888, HST-AR-13896, HST-AR-14282, HST-AR-14554, HST-GO-12914, and HST-GO-14191 from STScI.
DRW is supported by a fellowship from the Alfred P. Sloan Foundation.
AW was supported by NASA through grants HST-GO-14734 and HST-AR-15057 from STScI.
Support for PFH was provided by an Alfred P. Sloan Research Fellowship, NASA ATP Grant NNX14AH35G, and NSF Collaborative Research Grant \#1411920 and CAREER grant \#1455342.
DK and TKC were supported by NSF grants AST-1412153 and AST-1715101 and the Cottrell Scholar Award from the Research Corporation for Science Advancement.
CAFG was supported by NSF through grants AST-1412836, AST-1517491, AST-1715216, and CAREER award AST-1652522, and by NASA through grant NNX15AB22G.
We ran numerical calculations on the Caltech compute cluster ``Wheeler,'' allocations TG-AST130039 \& TG-AST150080 granted by the Extreme Science and Engineering Discovery Environment (XSEDE) supported by the NSF, and the NASA HEC Program through the NAS Division at Ames Research Center and the NCCS at Goddard Space Flight Center.
The analysis in this paper relied on the python packages \texttt{NumPy} \citep{vanderwalt_2011}, \texttt{Matplotlib} \citep{Hunter_2007}, and \texttt{AstroPy} \citep{Astropy_2013}.




\bibliographystyle{mnras}

\appendix
\section{Toy model for unresolved HI}
\label{sec:toy_model}

\begin{figure*}
\includegraphics[width=\textwidth]{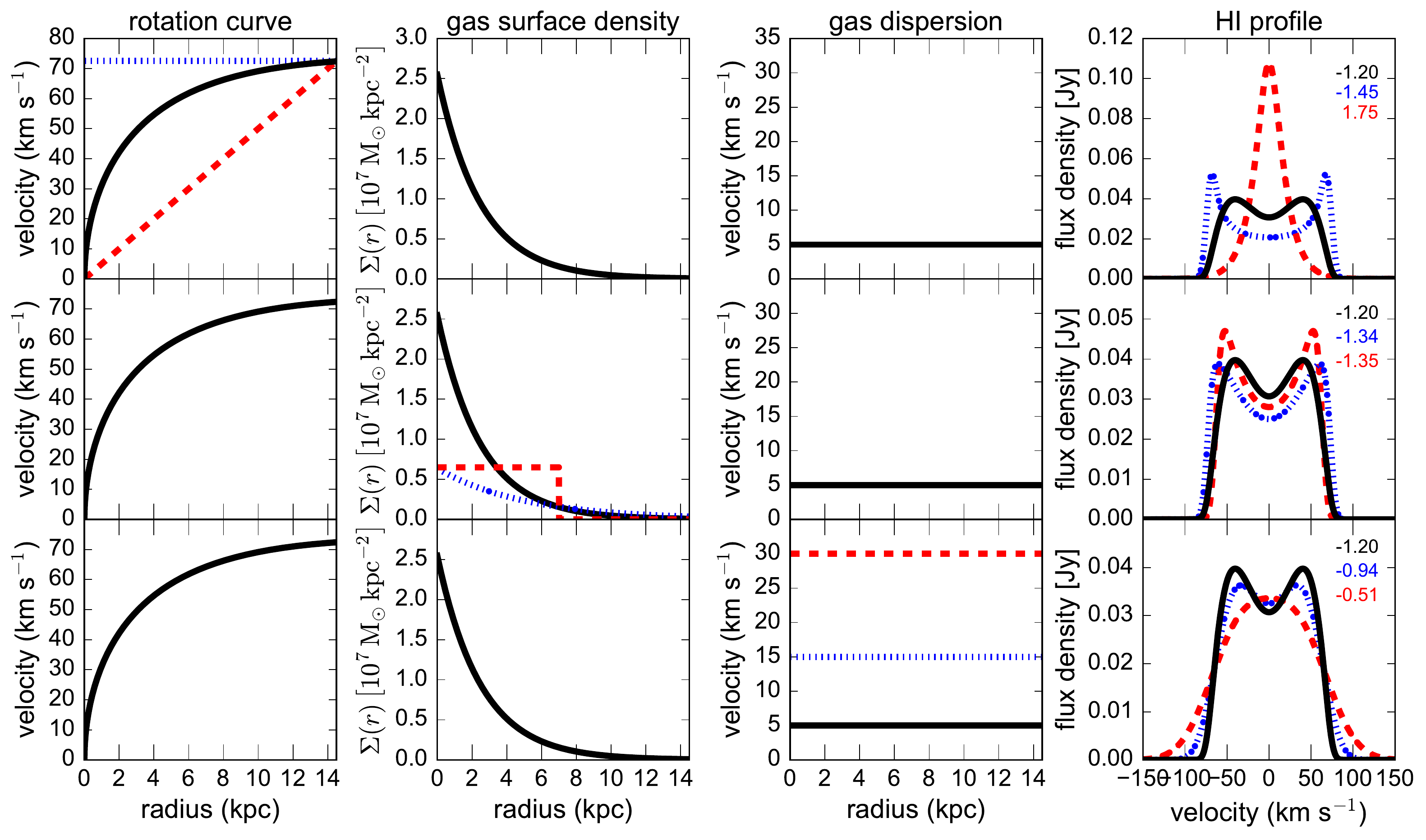}
\caption{Schematic effect of the rotation curve, gas surface density, and gas velocity dispersion on a galaxy's observable HI profile as predicted by our toy model (Section~\ref{sec:toy_model}). The black line shows the same fiducial model in all three rows. We vary the rotation curve (top), surface density profile (middle), and dispersion (bottom), using Equation~\ref{eqn:S_HI_toy} to predict the HI profile for each configuration. Numbers in the upper right panels indicate the profile's kurtosis. We assume $M_{\rm HI} = 10^9 M_{\odot}$ and a distance of 30 Mpc in all panels. Increasing the velocity dispersion or decreasing the rotation velocity in the region occupied by most of the gas causes the line profile to transition from its characteristic double-horned profile to a single-peaked profile with broader wings.}
\label{fig:toy_model}
\end{figure*}

In this section, we use a simple analytic model to show how the shape of a galaxy's integrated HI profile depends on its rotation curve, gas velocity dispersion, and gas surface density profile. We begin by considering a thin rotating gas disk in the $x-y$ plane, with HI surface density profile $\Sigma(r)$ and an intrinsic rotation curve $v_{\phi,0}(r)$, viewed at inclination angle $i$; for a rotation-supported, spherically symmetric system in equilibrium, $v_{\phi,0}(r) = v_c(r) = \sqrt{GM(<r)/r}$.\footnote{We generally treat the rotation curve as arising from a spherically symmetric mass distribution, but we emphasize that the model is agnostic to the precise form of the rotation curve; for example, accounting for the potential due to the disk simply requires substituting the appropriate form of $v_{c}(r)$ for $v_{\phi,0}(r)$.} We first consider an infinitesimal parcel of gas at $(r,\phi)$, defining the coordinate system such that $\phi = 0$ points toward the observer, who is at an angle $\theta = \pi/2-i$ above the disk midplane. The parcel's projected radial velocity along a line-of-sight to the observer is 
\begin{align}
\label{eqn:v_los}
v_{\rm los}(r,\phi)=v_{\phi,0}(r)\sin\phi\sin i=v_{\phi}(r)\sin\phi,
\end{align}
where we define $v_{\phi}(r) = v_{\phi,0}(r)\sin i$. The mass in this gas parcel is ${\rm d}m=\Sigma(r)\times{\rm d}r\times r{\rm d}\phi$, so the gas mass at radius $r$ in an infinitesimal velocity bin ${\rm d}v_{\rm los}$ follows 
\begin{align}
\label{eqn:dm_dv}
\frac{{\rm d}m}{{\rm d}v_{\rm los}}(r, v_{\rm los})=2\frac{{\rm d}m}{{\rm d}\phi}\frac{{\rm d}\phi}{{\rm d}v_{\rm los}}=2r\Sigma\left(r\right){\rm d}r\frac{{\rm d}\phi}{{\rm d}v_{\rm los}},
\end{align}
where the extra factor of 2  arises because material at $\phi$ has the same $v_{\rm los}$ as material at $\pi - \phi$. Differentiating Equation~\ref{eqn:v_los} with respect to $\phi$ and inverting the derivative, we find 
\begin{align}
\label{eqn:dphi_dv}
\frac{{\rm d}\phi}{{\rm d}v_{\rm los}}(r, v_{\rm los})=\frac{1}{v_{\phi}\left(r\right)\cos\phi}=\frac{1}{\sqrt{v_{\phi}^{2}\left(r\right)-v_{\rm los}^{2}}}.
\end{align}
The total HI mass in the disk per velocity interval ${\rm d}v_{\rm los}$ can then be found by integrating Equation~\ref{eqn:dm_dv} over all\footnote{In practice, the integrand is undefined for $v_{\phi}^2 < v_{\rm los}^2$, so we only integrate over $r_{\rm min} <r<r_{\rm max}$, where $r_{\rm min}$ and $r_{\rm max}$ bound the region where $v_{\phi}(r) > v_{\rm los}$. This makes sense physically, because annuli where $v_{\phi}(r) < v_{\rm los}$ clearly will not have any material moving at velocity $v_{\rm los}$.} radii: 
\begin{align}
\label{eqn:full_dM_dv}
M(v_{\rm los})=\frac{{\rm d}M}{{\rm d}v_{\rm los}}=2\int_{0}^{\infty}\frac{\Sigma(r)r\,{\rm d}r}{\sqrt{v_{\phi}^{2}(r)-v_{\rm los}^{2}}}.
\end{align}
Thus far, we have assumed that the projected rotation velocity $v_{\phi}(r)$ is the same for all particles in a given annulus, as will be the case if the gas is supported purely by rotation. However, if there is intrinsic dispersion in the rotation velocity at a given radius, this assumption will not hold. If we take the azimuthal velocity distribution in a given annulus to be a Gaussian centered on $v_{\phi}(r)$ with dispersion $\sigma(r)$, then Equation~\ref{eqn:dphi_dv} for a single annulus must be convolved with a Gaussian. Again integrating over all radii, we obtain  
\begin{align}
\label{eqn:with_disp}
M(v_{{\rm los}})=\sqrt{\frac{2}{\pi}}\int_{0}^{\infty}\int_{-\infty}^{\infty}\frac{\exp\left[-\frac{\left(v_{{\rm los}}-\tilde{v}\right)^{2}}{2\sigma^{2}(r)}\right]\Sigma(r)r}{\sigma(r)\sqrt{v_{\phi}^{2}(r)-\tilde{v}^{2}}}\,{\rm d}\tilde{v}\,{\rm d}r,
\end{align}
which obeys the expected normalization condition that $M_{{\rm HI}}=\int_{-\infty}^{\infty}M\left(v_{{\rm los}}\right)\,{\rm d}v_{{\rm los}}=2\pi\int_{0}^{r}\Sigma\left(r\right)r\,{\rm d}r$.
Assuming negligible HI self-absorption, the observable 21-cm flux density is then 
\begin{align}
\label{eqn:S_HI_toy}
S_{{\rm 21}}(v_{{\rm los}})=4.24\times10^{-6}\left(\frac{M(v_{{\rm los}})}{M_{\odot}}\right)\left(\frac{D}{{\rm Mpc}}\right)^{-2}\,{\rm Jy}.
\end{align}
We note that a very similar model was used by \citet{Obreschkow_2009} to predict the unresolved HI and CO lines of galaxies for their semi-analytic model.

In Figure~\ref{fig:toy_model}, we show how the HI profile shapes predicted by Equation~\ref{eqn:with_disp} depend on $v_{\phi}(r)$ (top), $\Sigma(r)$ (middle) $\sigma(r)$ (bottom). In each row, the black line shows the same fiducial model: a \citet{Hernquist_1990} rotation curve with total mass $M = 10^{11}M_{\odot}$ and scale length $a=20$ kpc, an exponential HI surface density with scale length $r_d = 2.5$ kpc, and a constant gas dispersion of $\sigma = 5\,\rm km\,s^{-1}$. We then show the effect of varying each of these parameters, considering constant and linearly rising rotation curves (top row), an exponential surface density with $r_d = 5$ kpc, and a constant density disk (middle row), and gas dispersions of $\sigma = 15$ and 30 $\rm km\,s^{-1}$ (bottom row). 

Figure~\ref{fig:toy_model} shows that double-horned HI profiles arise naturally for rotating geometries when the gas velocity dispersion is low compared to the maximum rotation velocity and a significant fraction of the gas mass ($M \sim r^2\Sigma(r)$) is rotating at near the maximum rotation velocity. A single-peaked HI profile with high kurtosis and shallow wings can be produced either by a high velocity dispersion (dashed red line, bottom row) \textit{or} by a low rotation velocity in the region where most of the gas is (dashed red line, top row). Low-mass galaxies are thus not guaranteed to produce double-peaked HI profiles even if their gas is rotation-supported; halos with solid-body rotation curves extending to the outskirts of the HI disk will generically produce single-peaked unresolved profiles.

For the simulated galaxies, we can measure $v_c(r)$ and $v_{\phi}(r)$ directly.\footnote{These curves are presented for all the simulations analyzed in this work in the Appendix of \citetalias{ElBadry_2018}.} In most of the simulated galaxies with $M_{\rm star} \lesssim 10^{10}M_{\odot}$, the circular velocity $v_c(r)$ is qualitatively similar to the default rotation curve adopted in Figure~\ref{fig:toy_model}, but the true gas rotation velocity $v_{\phi}(r)$ is significantly lower; for these systems, the lack of double-peaked profiles is verifiably due to enhanced dispersion support and non-circular motions, not a linearly-rising circular velocity curve.

For the simple model considered here, double-peaked line profiles can only be produced by rotation. The model will break down, however, when dispersion support begins to dominate or outflows drive large non-circular motions. In such cases, it is in principle possible to produce a double-horned profile purely due to radial kinematics; i.e., when a galaxy with a large bipolar outflow containing most of the HI mass is viewed head on. Such outflows are common in the simulated low-mass galaxies we study \citep{Muratov_2015, ElBadry_2016, ElBadry_2018}, but the simulated galaxies that are not rotation-supported generally do not produce double-peaked lines. This, coupled with the ubiquity of double-peaked profiles in the observed sample, suggests that rotation is the primary driver of the double-peaked profiles and lower kurtosis values found in the observed sample.

Several resolved observational studies \citep{Swaters_2009, Oh_2015, Iorio_2017} have found some low-mass galaxies to have solid-body rotation curves out to $\sim$2 disk scale radii; as illustrated in the top row of Figure~\ref{fig:toy_model}, such systems can produce single-peaked unresolved profiles even if they have rotationally supported disks. We therefore caution that the increase in the kurtosis values of the \textit{observed} sample at low masses could be attributed either to increased dispersion support or to linearly rising rotation curves. 

\begin{figure}
\includegraphics[width=\columnwidth]{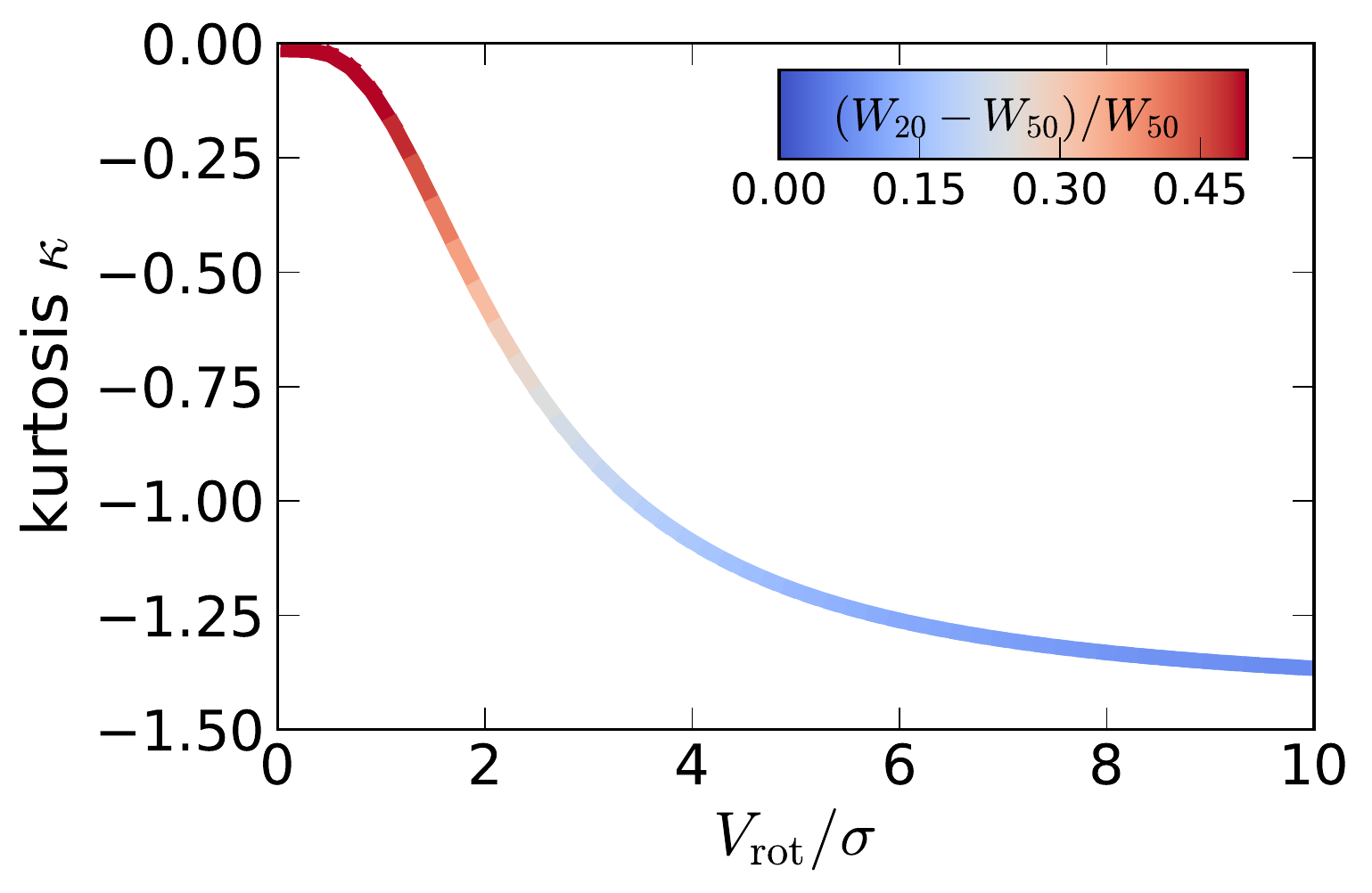}
\caption{Relation between HI line kurtosis, $(W_{20} - W_{50})/W_{50}$, and $V_{\rm rot}/\sigma$ as predicted by our toy model, assuming a Hernquist rotation curve and exponential disk surface density profile (see Section~\ref{sec:toy_model}). Kurtosis and $(W_{20} - W_{50})/W_{50}$ are both measures of the steepness of the unresolved HI profile's wings and are inversely correlated with $V_{\rm rot}/\sigma$.}
\label{fig:toy_model_kurt}
\end{figure}

In Figure~\ref{fig:toy_model_kurt}, we show the relation predicted by this analytic model between $V_{\rm rot}/\sigma$, unresolved line kurtosis, and the steepness of the line's wings. Here we assume a Hernquist rotation curve, exponential surface density profile, and constant gas dispersion, but our predictions are qualitatively similar for a wide range of models. We define $V_{\rm rot}$ as the maximum value of $v_{\phi}(r)$; we note that $V_{\rm rot}/\sigma$ defined in this way will generally be higher than the observationally-motivated $V/\sigma$ defined in Equation~\ref{eqn:v_sig} due to projection effects. In agreement with the trend found for our simulated galaxies (Figure~\ref{fig:vsig}), the model predicts that $\kappa$ and $(W_{20}-W_{50})/W_{50}$ increase monotonically as galaxies transition from rotational to dispersion support.

Although we have not done so in this work, we note that this model can also be applied to fit the observed HI profiles of real galaxies; i.e., given parameterized models for $v_{\phi}(r)$, $\sigma(r)$, and $\Sigma(r)$, it is a straightforward exercise in optimization or sampling to determine what set of parameters produce an unresolved profile in best agreement with an observed profile. In this case, the tight correlation between observed galaxies' HI disk sizes and HI masses \citep[e.g.][]{Wang_2015} places a prior on $\Sigma(r)$, so that fitting an observed 21-cm profile strongly constrains $v_{\phi}(r)$ and $\sigma(r)$. Previous works \citep[e.g.][]{Westmeier_2014, Stewart_2014} have presented other flexible fitting functions for unresolved 21-cm profiles that capture most of the observed diversity in line shapes. The advantage of using a model such as the one introduced here is that the parameters returned from fitting are directly interpretable in terms of physical properties of the galaxy. The realization of next-generation radio facilities such as WALLABY and the SKA will produce HI observations of more than half a million galaxies, the vast majority of which will be spatially unresolved or marginally resolved  \citep{Dewdney_2009, Duffy_2012}. Interpreting these galaxies' HI profiles with physically motivated models will place strong statistical constraints on the distribution and kinematics of gas throughout the galaxy population. 

\section{Relation between morphology, star formation history, and cores}
\label{sec:sfhs}
\begin{figure}
\includegraphics[width=\columnwidth]{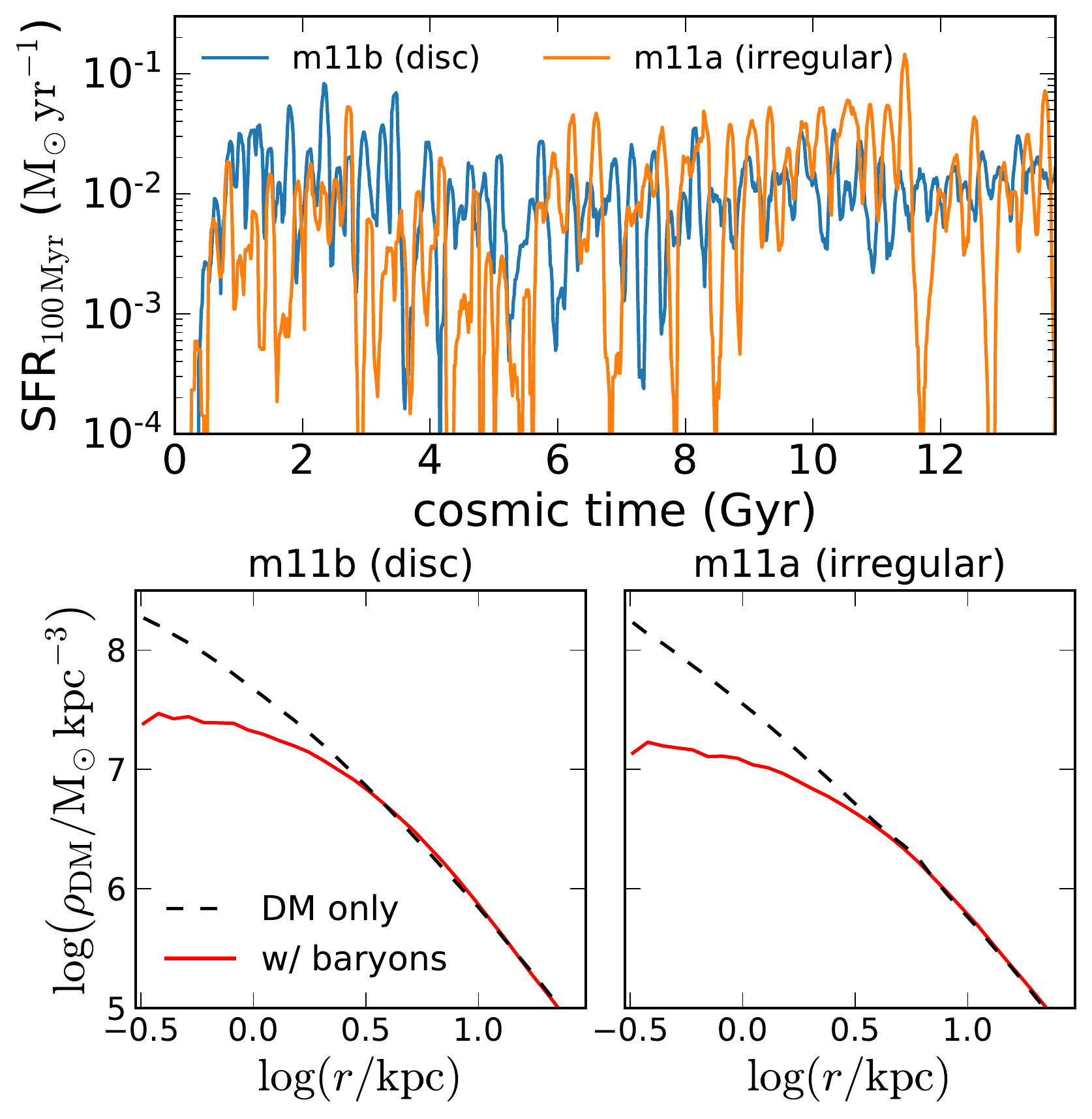}
\caption{\textbf{Top}: Star formation histories of two of our simulated galaxies with similar stellar and halo masses but different morphologies. The disky galaxy has a substantially less bursty SFH at late times than the irregular galaxy. \textbf{Bottom}: Dark matter density profiles at $z=0$ of the two halos from our standard simulations with baryons and feedback compared to profiles of the same halos in dark matter only simulations. Both galaxies have a cored density profile, but the irregular galaxy with a burstier SFH forms a somewhat larger core.}
\label{fig:m11b_vs_m11a}
\end{figure}

As discussed in Section~\ref{sec:conclusion}, the dispersion-supported kinematics and irregular morphologies of many of our simulated low-mass galaxies are likely related to their bursty star formation histories. To investigate this further, we compare in Figure~\ref{fig:m11b_vs_m11a} the SFHs\footnote{Here the SFH is defined ``archaeologically'' for all stars within 20 kpc at $z=0$, and the SFR at a given time represents the average over the previous 100 Myr. The SFH defined in this way differs slightly from that of the main progenitor, since a small fraction (< 10\%) of the stars in the galaxies at $z=0$ formed ex situ.} of \texttt{m11a} and \texttt{m11b}, the ``irregular'' and ``disky'' galaxies shown in Figure~\ref{fig:schematic}. Both galaxies have bursty star formation histories at early times, but the SFH of \texttt{m11b} becomes calmer at late times, while that of \texttt{m11a} remains quite bursty until $z=0$. In \texttt{m11a}, strong bursts of star formation and subsequent temporary quenched episodes occur when a large fraction of the galaxy's gas falls into the galactic center and accumulates at high densities. Because most of the gas in \texttt{m11b} is angular momentum-supported at late times, it cannot coherently fall into the center, and star formation is regulated more locally in the disk; see \citetalias{ElBadry_2018} for details. 

In the bottom panels of Figure~\ref{fig:m11b_vs_m11a}, we show the $z=0$ dark matter density profiles of the two halos. To highlight the effects of feedback-driven potential fluctuations in reducing the central density of dark matter, we compare the DM density profiles from the main simulations analyzed here to those from collisionless dark matter only simulations of the same zoom-in region.\footnote{We multiply the density profile from the DM-only simulation by $\left(1-\Omega_{{\rm b}}/\Omega_{{\rm m}}\right)$ to correct for the lack of baryons.} Consistent with previous work on coring at this mass scale \citep[e.g.][]{Chan_2015}, we find that both galaxies have substantially reduced DM densities in the central $\sim$few kpc in the simulations with baryons, indicating that feedback-driven potential fluctuations have redistributed DM from the central regions to larger radii. 

The core is somewhat larger in \texttt{m11a}, the dispersion-supported galaxy. This is not surprising, since the galaxy's burstier SFH at late times indicates that it has undergone larger outflows and rapid potential fluctuations at late times (see e.g. \citealt{Pontzen_2014} for discussion of the conditions required for feedback-driven core formation). The presence of a core in \texttt{m11b} can likely be attributed to its bursty SFH at early times \citep[e.g.][]{Madau_2014, Onorbe_2015}. Indeed, we find that the two galaxies have qualitatively similar morphologies and evolutionary histories until $z\approx 0.5$, when \texttt{m11b} goes through a merger. This brings in a large quantity of high-angular momentum gas, and the galaxy quickly forms a rotation-supported disk thereafter. 

\bsp	
\label{lastpage}
\end{document}